\documentclass[12pt]{article}
\newcommand {\citeAY} [1] {\citeNP {#1}}%
\newcommand {\citeAPY}[1] {\citeN  {#1}}%
\usepackage{xchicago}
\renewcommand {\showoriginalref}[1]{}
\renewcommand {\showCODEN}[1]{}
\renewcommand {\showISSN}[1]{}
\renewcommand {\showMR}[3]{}

\newcommand\eq[1] {(\ref{#1})}

\newcommand\labfig[1] {\label{fig:#1}}

\newcommand{\bfm}[1]{\mbox{\boldmath ${#1}$}}
\newcommand{\nonum}{\nonumber \\}
\newcommand{\beqa}{\begin{eqnarray}}
\newcommand{\eeqa}[1]{\label{#1}\end{eqnarray}}
\newcommand{\beq}{\begin{equation}}
\newcommand{\eeq}[1]{\label{#1}\end{equation}}
\newcommand{\R}{\mbox{R}}

\newcommand{\Grad}{\nabla}
\newcommand{\Div}{\nabla \cdot}
\newcommand{\Curl}{\nabla \times}

\newcommand{\Md}{\partial}

\newcommand{\Ga}{\alpha}
\newcommand{\Gb}{\beta}
\newcommand{\Gd}{\delta}

\newcommand{\Gve}{\varepsilon}

\newcommand{\Gg}{\gamma}

\newcommand{\Gk}{\kappa}

\newcommand{\Gn}{\eta}
\newcommand{\Gm}{\mu}

\newcommand{\Gt}{\theta}

\newcommand{\Gr}{\rho}

\newcommand{\Gj}{\tau}

\newcommand{\Go}{\omega}

\newcommand{\GD}{\Delta}

\newcommand{\BGve}{\bfm\varepsilon}

\def\Bu{{\bf u}}
\def\Bv{{\bf v}}

\def\Bx{{\bf x}}
\def\By{{\bf y}}

\def\BA{{\bf A}}
\def\BB{{\bf B}}

\def\BD{{\bf D}}
\def\BE{{\bf E}}

\def\BH{{\bf H}}
\def\BI{{\bf I}}

\def\BM{{\bf M}}

\def\BU{{\bf U}}
\def\BV{{\bf V}}

\usepackage{graphics}
\usepackage{amssymb}
\begin{document}
\vspace{-1in}
\title{Solutions in folded geometries, and associated cloaking due to anomalous resonance}
\author{Graeme W. Milton\\
\small{Department of Mathematics, University of Utah, Salt Lake City UT 84112, USA}\\ \\
Nicolae-Alexandru P. Nicorovici and Ross C. McPhedran\\
\small{ARC Center of Excellence for Ultrahigh-bandwidth Devices for Optical Systems (CUDOS)}, \\
\small{School of Physics, University of Sydney, Sydney NSW 2006, Australia}\\ \\
Kirill Cherednichenko\\
\small{School of Mathematics, Cardiff University, Senghennydd Road},
\\ {\small Cardiff, CF24 4AG, United Kingdom}\\ \\
Zubin Jacob\\
\small{Birck Nanotechnology Center, Department of Electrical and Computer
 Engineering}, \\ \small{Purdue University, West Lafayette IN 47907, USA}}

\date{}
\maketitle
\begin{abstract}
Solutions for the fields in a coated cylinder where the core radius is bigger than the shell radius
are seemingly unphysical, but can be given a physical meaning if one transforms to an equivalent
problem by unfolding the geometry. In particular the unfolded material can act as an impedance matched
hyperlens, and as the loss in the lens goes to zero finite collections
of polarizable line dipoles lying within
a critical region surrounding the hyperlens are shown to be cloaked having vanishingly small dipole moments.
This cloaking, which occurs both in the folded geometry and the equivalent unfolded one, is due to
anomalous resonance, where the collection of dipoles generates an anomalously resonant field, which
acts back on the dipoles to essentially cancel the external fields acting on them.

\vskip2mm

\end{abstract}

\section{Introduction}
\setcounter{equation}{0}

Analytical solutions have played an important role in understanding the
electromagnetic response of inclusions to an applied field.
In these analytic solutions nothing prevents one from substituting seemingly
unphysical values of the parameters. For example, for a coated spherical
inclusion with core radius $r_{\rm c}$ and shell radius $r_{\rm s}$ one may substitute
into the analytic solution for the fields parameter values $r_{\rm c}$ and $r_{\rm s}$ with $r_{\rm c}>r_{\rm s}$. Is there any physical significance
to such solutions? 
Introducing the novel concept (from the viewpoint
of classical electromagnetism) of folded geometries and
building upon the ideas of \citeAPY{Leonhardt:2006:GRE}
let us first show that ``yes there is''. 

Specifically, for simplicity, we analyze in the quasistatic limit
the transverse magnetic (TM) solution for a coated cylindrical
inclusion. In the usual situation, it is filled with an isotropic core material
having a homogeneous complex dielectric constant $\Gve_{\rm c}$ and radius $r_{\rm c}$, embedded
in a homogeneous isotropic shell of dielectric constant $\Gve_{\rm s}$ having radii $r_{\rm c}$ and
$r_{\rm s}$, with $r_{\rm s}>r_{\rm c}$, which itself is embedded in a homogeneous isotropic matrix having
dielectric constant $\Gve_{\rm m}$. The potential $V$ takes values $V_{\rm c}$,
$V_{\rm s}$ and $V_{\rm m}$ in the core, shell, and matrix respectively. Each of
these are harmonic functions (satisfying $\GD V=0$) within their
respective domains, except at singularities which we assume are
confined to a finite set of points in the matrix. At the interfaces
they satisfy the boundary  conditions \beqa V_{\rm c}|_{r=r_{\rm c}}& = &
V_{\rm s}|_{r=r_{\rm c}},\quad\quad V_{\rm s}|_{r=r_{\rm s}}=V_{\rm m}|_{r=r_{\rm s}} \nonum
\Gve_{\rm c}\frac{\Md V_{\rm c}}{\Md r}\bigg|_{r=r_{\rm c}}& = &  \Gve_{\rm s}\frac{\Md
V_{\rm s}}{\Md r}\bigg|_{r=r_{\rm c}},\quad\quad \Gve_{\rm s}\frac{\Md V_{\rm s}}{\Md
r}\bigg|_{r=r_{\rm s}}=\Gve_{\rm m}\frac{\Md V_{\rm m}}{\Md r}\bigg|_{r=r_{\rm s}}.
\eeqa{0.1} These equations still make mathematical sense if
$r_{\rm c}>r_{\rm s}$: we look for harmonic potentials $V_{\rm c}$, $V_{\rm s}$ and $V_{\rm m}$
defined in the respective regions $r\leq r_{\rm c}$, $r_{\rm s}\leq r \leq r_{\rm c}$
and $r>r_{\rm s}$, and satisfying the boundary conditions \eq{0.1}, where
now $\Gve_{\rm c}$, $\Gve_{\rm s}$ and $\Gve_{\rm m}$ are regarded as mathematical
parameters entering these boundary conditions. The dielectric tensor
field $\BGve(\Bx)$ takes values \beqa \BGve(\Bx) & = & \Gve_{\rm c}\BI
\quad {\rm in~the~core}, \nonum
                & = & -\Gve_{\rm s}\BI \quad {\rm in~the~shell}, \nonum
                & = & \Gve_{\rm m}\BI \quad {\rm in~the~matrix}, \nonum
\eeqa{0.2}
 with the choices of sign here being motivated by the effect of folding of space "back on itself", which
 affects the direction of derivatives. Indeed, flux will be conserved only if the radial
component of the displacement field $\BD(\Bx)=-\BGve(\Bx)\Grad V$ changes
sign, but maintains magnitude, at these interfaces: if $\Div \BD=0$ then one can draw a flow field for $\BD$ with arrows
and (by conservation of flux) the arrows must reverse direction at the
interface. The interface
conditions \eq{0.1} are compatible with this constraint provided
$\BGve(\Bx)$ is given by \eq{0.2}.

To make physical sense of such a solution we recall the fact that the quasistatic equations (and more generally,
the equations of electromagnetism) retain their form under coordinate transformations. Specifically if $V(\Bx)$
is a solution to
\beq \Div\BGve(\Bx)\Grad V(\Bx)=0 \eeq{0.3}
and $\Bx'(\Bx)$ is a transformation to a new curvilinear coordinate system, then the potential
$V'(\Bx')\equiv V(\Bx(\Bx'))$, where $\Bx(\Bx')$ is the inverse transformation, satisfies
\beq  \Grad'\cdot\BGve'(\Bx')\Grad' V'(\Bx')=0 \eeq{0.4}
where the dielectric tensor, viewed as a contravariant tensor density, has been transformed
according to the standard formula
\beq \BGve'(\Bx')=|{\rm det}\BA(\Bx)|^{-1}\BA(\Bx)\BGve(\Bx)\BA^T(\Bx) \eeq{0.5}
in which $\BA=\Grad\Bx'(\Bx)$ is the Jacobian, and $\Bx=\Bx(\Bx')$. The equation \eq{0.4} can
be reinterpreted as a quasistatic equation in a body with dielectric constant $\BGve'(\Bx')$
in which $\Bx'=(x'_1,x'_2,x'_3)$ are now regarded as Cartesian coordinates. The displacement field and the electric field
$\BE(\Bx)=-\Grad V(\Bx)$ transform to
\beq
\BD'(\Bx')=|{\rm det}\BA(\Bx)|^{-1}\BA(\Bx) \BD(\Bx),~~\BE'(\Bx')=[\BA^T(\Bx)]^{-1} \BE(\Bx).
\eeq{0.5a}

To turn the unphysical solution in the folded geometry, with $r_{\rm c}>r_{\rm s}$, into a physical solution
we use a coordinate transformation which unfolds the geometry. Consider the standard polar coordinates $(r,\Gt)$
and $(r',\Gt')$ in the folded and transformed geometries respectively. Then the simplest unfolding mapping, as sketched in Fig.1,  is
given by $\Gt'=\Gt$ and
\beqa r'& = & r_{\rm c}^{-1}[r_{\rm s}-a(r_{\rm c}-r_{\rm s})]r, \quad {\rm in~the~core}, \nonum
                & = & r_{\rm s}-a(r-r_{\rm s}) \quad {\rm in~the~shell}, \nonum
                & = & r \quad {\rm in~the~matrix}, \nonum
\eeqa{0.6}
where $a$ is a fixed positive constant less than $r_{\rm s}/(r_{\rm c}-r_{\rm s})$. We emphasize that the pair $(r,\Gt)$ 
with $r_c>r>r_s$ and $2\pi>\Gt\geq 0$ does not suffice to uniquely specify a point in the folded geometry: one has to specify
whether the point lies in the core, shell, or matrix. In a folded geometry it is as if space overlaps itself but without
intersection: as one goes continuously on a straight line trajectory from the origin, first one moves in the core and 
the radius increases until one encounters the core radius $r_c$ , then one moves into the shell and the radius decreases 
until one reaches the shell radius $r_s<r_c$, where one moves into the matrix and the radius increases again.   
With this definition, the unfolding mapping \eq{0.6} is globally a 1 to 1 mapping. 

It is clear from \eq{0.6} that $r'_{\rm s}=r_{\rm s}>r_{\rm c}'=r_{\rm s}-a(r_{\rm c}-r_{\rm s})$. The
inverse folding transformation $\Bx(\Bx')$ takes the same form as \eq{0.6} with $r_{\rm c}$, $r_{\rm s}$ and $a$ replaced
by $r'_{\rm c}$, $r'_{\rm s}$ and $a^{-1}$ respectively, and the roles of $r$ and $r'$ swapped. Using the expression
\eq{0.5} and the formula for the unfolding map, which in particular implies that in the shell
\beq \Bx'=-a\Bx+b\Bx/\sqrt{\Bx\cdot\Bx}, \quad
\BA=\Grad\Bx'=(b/r-a)\BI-b\Bx\otimes\Bx/r^3,
\eeq{0.6a}
where $b=(1+a)r_{\rm s}$,
we get expressions for the dielectric tensor in the core, shell
and matrix in the unfolded geometry
\beq \BGve_{\rm c}'=\Gve_{\rm c}\BI,\quad
     \BGve_{\rm s}'=-\frac{(b/r-a)^2\BI+(2ab/r^3-b^2/r^4)\Bx\otimes\Bx}{a(b/r-a)}\Gve_{\rm s},
\quad \BGve_{\rm m}'=\Gve_{\rm m}\BI.
\eeq{0.7}
To be physically realizable we require that $\BGve_{\rm c}'$, $\BGve_{\rm s}'$ and $\BGve_{\rm m}'$ have positive semi-definite
imaginary parts, which requires that $\Gve_{\rm c}$ and $\Gve_{\rm m}$ have a non-negative imaginary part, while $\Gve_{\rm s}$
has a non-positive imaginary part (as can be seen directly from \eq{0.2} and \eq{0.5}).
In summary we see that seemingly paradoxical
geometries may be transformed into a physically comprehensible form,
which may prove an interesting direction for future research.

\begin{figure}
\vspace{2in}
\hspace{-2in}
\centering
{\resizebox{1.5in}{1.5in}
{\includegraphics[0in,0in][3in,3in]{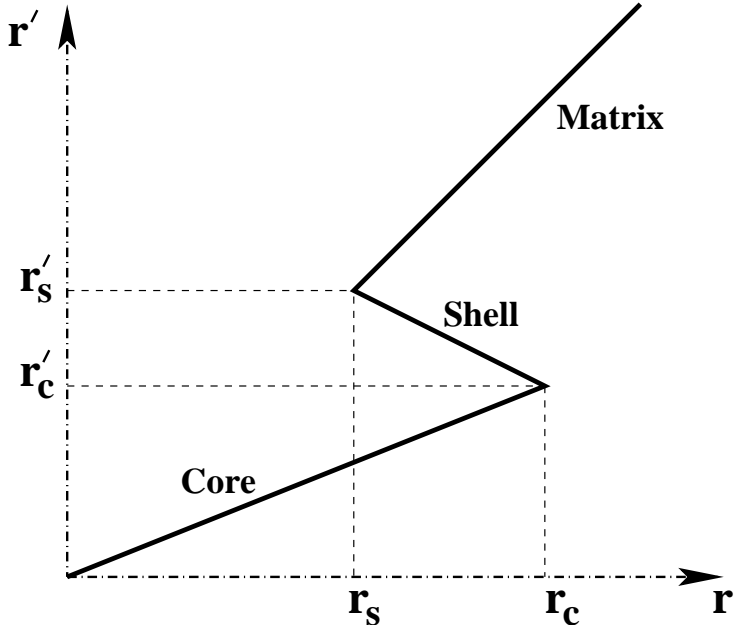}}}
\hspace{2.0in}
\vspace{0.1in}
\caption{Sketch of the unfolding transformation \eq{0.6}, where $r$ and $r'$ are the radial coordinates in the
folded and unfolded geometries. Note that $r_s'=r_s$ since the mapping is the identity map in the matrix. }
\labfig{-1}
\end{figure}

When $\Gve_{\rm s}=\Gve_{\rm m}$ the response of the coated cylinder in the folded geometry is equivalent
to that of a solid cylinder of radius $r_{\rm c}$
and dielectric constant $\Gve_{\rm c}$. The potential in the shell in the folded region between $r_{\rm c}$ and $r_{\rm s}$
is the same as that in the matrix in this region and is the analytic extension of the potential
surrounding the solid cylinder
provided there are no singularities in this analytic extension- otherwise a solution does not exist. So in
the unfolded geometry the shell with dielectric tensor $\BGve_{\rm s}'(\Bx')$ acts to magnify the core by
a factor of $r_{\rm c}/r_{\rm c}'$ so it responds like a solid cylinder of radius $r_{\rm c}$ and dielectric constant $\Gve_{\rm c}$.
We call such a shell an impedance-matched hyperlens lens in recognition of the pioneering work of \citeAPY{Kildishev:2007:IMH} who showed that it would magnify fixed sources in core, not just in the quasistatic limit, but also for the full
Helmholtz equation (provided the magnetic permeability was also suitably chosen). Such lenses
were first considered by \citeAPY{Rahm:2008:DEC} as electromagnetic concentrators. Although
both groups assumed $r_{\rm s}>r_{\rm c}$, their analysis extends directly to the case
$r_{\rm c}>r_{\rm s}$. Other hyperlenses with magnifying properties were
studied by \citeAPY{Jacob:2006:OHF} and \citeAPY{Salandrino:2006:FFS}.

This equivalence is similar to the result of \citeAPY{Nicorovici:1994:ODP} who found that a coated dielectric
cylinder with radii $r_{\rm s}>r_{\rm c}$ and moduli $\Gve_{\rm s}=-\Gve_{\rm m}$ would have the same quasistatic response as a
solid cylinder of radius $r_*=r_{\rm s}^2/r_{\rm c}$ and dielectric constant $\Gve_{\rm c}$, i.e. the shell,
of dielectric constant $\Gve_{\rm s}=-\Gve_{\rm m}$, now known as a cylindrical superlens, acts to magnify
the core by the factor $h=r_{\rm s}^2/r_{\rm c}^2$. This equivalence implied that a line source at radius $r_0>r_{\rm s}$
in the matrix would generate a potential which appeared like it originated from the line source
plus an image line source at the radius $r_*^2/r_0$ which would {\it be in the matrix} when
$r_*^2/r_0>r_{\rm s}$. They found that the actual potential in the matrix converged as $\Gve_{\rm s}\to -\Gve_{\rm m}$
to this singular potential at radii greater than $r_*^2/r_0$ and numerically found that the actual
potential developed large oscillations at smaller radii. (See, in particular, the sentence beginning
with ``These fluctuations become less pronounced..'' above figure 2 in that paper.) To our
knowledge this was the first discovery of an apparent (ghost) singularity in the field
surrounding an inclusion, or in effect the first example of perfect imaging (in quasistatics)
of a point or line source. The regions where the field diverges were later
called regions of anomalous resonance (\citeAY{Milton:2005:PSQ}).

In a subsequent development \citeAPY{Pendry:2000:NRM} made the bold claim
that the Veselago lens (\citeAY{Veselago:1967:ESS})
consisting of a slab of thickness $d$ with dielectric constant $\Gve_{\rm s}=-1$ and magnetic permittivity $\Gm_{\rm s}=-1$,
surrounded by a medium  with dielectric constant $\Gve_{\rm m}=1$ and magnetic permittivity $\Gm_{\rm m}=1$,
would behave as a superlens: a line source at a distance $d_0$ in front of the slab, would appear to
have an image line source at a distance $d-d_0$ behind the slab. When  $\Gve_{\rm s}$ and $\Gm_{\rm s}$
approached $-\Gve_{\rm m}$ and $-\Gm_{\rm m}$ (having a very small imaginary part) the actual fields
behind the slab converged to these singular fields behind the image,
but diverged between the image and the slab. There was also
a seeming paradox (pointed out to GWM by Alexei Efros): if the source was closer than a distance
$d/2$ to the lens then the electromagnetic power dissipated in the lens per unit time by a
constant amplitude source would approach
infinity as the loss went to zero. This paradox was resolved by \citeAPY{Milton:2005:PSQ}
who showed that when $d_0<d/2$ then
the anomalously resonant fields acting on the source act as a sort of ``optical molasses'' against which the
source has to do a tremendous amount of work to maintain its amplitude. Subsequently it was
found that a polarizable dipolar line source or single constant energy line source
becomes ``cloaked''
if it is within a distance $d/2$ of the slab lens or within a radius $\sqrt{r_{\rm s}^3/r_{\rm c}}$ of
a cylindrical superlens (with the core having dielectric constant $\Gve_{\rm c}=\Gve_{\rm m}$). Its dipole
moment, and consequently its effect on the field outside a certain
distance from the lens, becomes vanishingly small.
The energy generated by a constant energy source, like the energy generated by two opposing sources
on opposite sides of a slab lens (\citeAY{Cui:2005:LEE}; \citeAY{Boardman:2006:NRC}) is effectively trapped within the cloaking region.
This cloaking
was proved (\citeAY{Milton:2006:CEA}) and numerically verified (\citeAY{Nicorovici:2007:OCT}) to extend to
collections of finitely many
polarizable dipoles. Also arguments were presented (\citeAY{Milton:2006:OPL})
which suggested that a line dipole which
was ``switched on'' at time $t=0$ in front of a perfect lens with no loss, having $\Gve_{\rm s}=-\Gve_{\rm m}$
and $\Gm_{\rm s}=-\Gm_{\rm m}$, would become cloaked in the limit $t\to \infty$. On the other hand
\citeAPY{Bruno:2007:SCS}
showed that a dielectric body such as a solid cylinder of finite radius in the cloaking region
would only be partially
and not fully cloaked in the limit as the loss goes to zero. One can conclude that
a dielectric body {\it is neither perfectly cloaked nor perfectly imaged by superlenses}
(in the limit as the loss goes to zero)
if it lies within the cloaking region.

Here we show that anomalous resonance and cloaking extends to folded cylindrical geometries, and therefore also
to the equivalent unfolded cylindrical geometries. This is not too surprising. \citeAPY{Leonhardt:2006:GRE}
realized that
the solution for the electromagnetic fields in the slab superlens can be viewed as the result of an
unfolding of space, and we know that anomalous resonance and cloaking are associated with superlenses.

There are important conceptual differences between the work of 
\citeAPY{Leonhardt:2006:GRE}, and our work. In their work the unfolding transformation
is applied to empty space, so that in the appropriate region one point gets 
mapped to three points, and a field at that point gets mapped to three fields.
In this context it is correct, as they do, to take transformations
of the moduli of the form \eq{0.5}, but without the absolute value around the Jacobian
of the determinant. In our approach, applied to the idealized ``perfect'' superlens with
the full-Maxwell equations, the unfolding 
transformation is applied to a folded geometry, and there is globally a one to one correspondence
between points in the folded geometry and the unfolded geometry. (The value of $\Bx$
in the folded geometry is not necessarily sufficient to specify a point: 
one also has to specify the manifold on which the point lies.) Given empty space
one first inserts a fold.
In the half of the fold that gets mapped
to the lens $\Curl$ gets replaced by $-\Curl$ 
in the Maxwell equations because of the change in handedness
of the space and 
the moduli are negative to ensure that the Maxwell 
equations $\Div\BD=\Div\BB=0$ remain satisfied in any source free region in the folded geometry. At a given
value of $\Bx$ in the fold the electromagnetic fields take the values $(\BE,\BD,\BH,\BB)$, $(\BE,-\BD,\BH,-\BB)$, 
and $(\BE,\BD,\BH, \BB)$ on the three different manifolds, 
where $\BE$, $\BD$, $\BH$ and $\BB$ are the electromagnetic fields
at $\Bx$ in the original empty space.
Thus the total displacement field density at $\Bx$ is $\BD$ (and not $3\BD$).
When transforming the moduli absolute
values around the Jacobian of the determinant are needed to ensure that Maxwell's
equations remain satisfied in the unfolded ``perfect'' superlens geometry. Our
introduction of folded geometries greatly enlarges the class of geometries to
which one can transform to simplify the analysis of a problem. This simplification
is analogous to the way one uses conformal transformations to map to a simpler 
problem.

For simplicity our analysis [which for the most part only requires minor modifications of
the analysis of \citeAPY{Milton:2006:CEA}]
is for two-dimensional quasistatics. Presumably analogous results hold for
the full (time harmonic) Maxwell equations in three dimensional folded spherical geometries, although
we have not explored this. Throughout the paper we use the symbol $\equiv$ to mean equal
by definition, and the symbol $\approx$ to mean approximately equal to.

\section{The Green function for a monopole and solutions for a dipole in the matrix}
\setcounter{equation}{0}
Let us consider the Green function $V(\Bx)$ for a point source (monopole) located in the matrix. Although unphysical
(because the net charge associated with the singularity oscillates in time) it is mathematically well defined,
and useful for deriving the potential associated with a dipole. This potential,
by definition, takes values $V_{\rm c}$, $V_{\rm s}$ and $V_{\rm m}$ in the core, shell, and matrix which
satisfy
\beq \GD V_{\rm c}=0, \quad \GD V_{\rm s}=0, \quad \GD V_{\rm m}=-\Gd(\Bx-\Bx_0)
\eeq{0.8}
in their respective domains, together with the boundary conditions \eq{0.1}, where $\Gd(\By)$ is the
standard Dirac delta function for a source located at $\By=0$. The problem of finding $V(\Bx)$ can be solved
explicitly using power series with respect to the complex coordinate $z=x_1+ {\rm i}x_2$, as follows. Note that
the Green function for the Laplace equation in  $\R^2$ is given by the formula
\beq V_0=-\frac{1}{4\pi}\bigl(\log(z-z_0)+\log(\overline{z}-\overline{z_0})\bigr)
=-\frac{1}{4\pi}\Bigg[(2\log| z_0|-\sum_{n=1}^\infty n^{-1}
\biggl(\frac{z}{z_0}\biggr)^n-\sum_{n=1}^\infty n^{-1}
\biggl(\frac{\overline{z}}{\overline{z_0}}\biggr)^n\Biggr].
\eeq{0.9}
This is the potential of a point monopole in a homogeneous free space.

We are looking for a solution $V_{\rm s,c,m}$ to the above problem (\eq{0.8} and \eq{0.1}) in the form of a
power series in each of the three regions:
\beqa
V_{\rm c}& = & \sum_{n=0}^\infty A_n^{({\rm c})} z^n+\sum_{n=0}^\infty B_n^{({\rm c})}\overline{z}^n, \nonum
V_{\rm s}& = & \sum_{n=-\infty}^\infty A_n^{({\rm s})} z^n+\sum_{n=-\infty}^\infty B_n^{({\rm s})}\overline{z}^n, \nonum
V_{\rm m}& = & V_0+\sum_{n=1}^{\infty} A_n^{({\rm m})} z^{-n}+\sum_{n=1}^{\infty} B_n^{({\rm m})}\overline{z}^{-n}
\eeqa{0.10}
The substitution of these series in the interface conditions \eq{0.1} yields via the identity
$r\partial/(\partial r)=z\partial/(\partial z)+\overline{z}\partial/(\partial\overline{z})$
explicit  expressions for the
coefficients $A^{({\rm c,s,m})}_n,$ $B^{({\rm c,s,m})}_n$. The formulae for
$V_{\rm c,s,m}$ can then be found, and are as follows:
\beqa V_{\rm c} & = & -\frac{1}{2\pi}\log|z_0|
+\frac{\Gve_{\rm s}\Gve_{\rm m}}{\pi(\Gve_{\rm s}-\Gve_{\rm c})(\Gve_{\rm m}-\Gve_{\rm s})}
\sum_{n=1}^{\infty}\left[\left(\frac{r_{\rm c}}{r_{\rm s}}\right)^{2n}+\Gd e^{i\phi}\right]^{-1}
\frac{1}{n}\left[\left(\frac{z}{z_0}\right)^n+\left(\frac{\overline{z}}{\overline{z_0}}\right)^n\right],
\nonum
V_{\rm s}& = & -\frac{1}{2\pi}\log|z_0| \nonum
& + & \frac{\Gve_{\rm m}}{2\pi\Gn_{\rm sc}(\Gve_{\rm m}-\Gve_{\rm s})}\sum_{n=1}^\infty\left[\left(\frac{r_{\rm c}}{r_{\rm s}}\right)^{2n}+\Gd e^{i\phi}\right]^{-1}
\frac{1}{n}\left[\left(\frac{z}{z_0}\right)^n+\left(\frac{\overline{z}}{\overline{z_0}}\right)^n\right] \nonum
& + & \frac{\Gve_{\rm m}}{2\pi(\Gve_{\rm m}-\Gve_{\rm s})}\sum_{n=1}^\infty\left[\left(\frac{r_{\rm c}}{r_{\rm s}}\right)^{2n}+\Gd e^{i\phi}\right]^{-1}
\frac{1}{n}\left[\left(\frac{z\overline{z_0}}{r_{\rm c}^2}\right)^{-n}+\left(\frac{\overline{z}z_0}{r_{\rm c}^2}\right)^{-n}\right]
\nonum
V_{\rm m} & = & V_0+\frac{1}{4\pi}\sum_{n=1}^\infty\left[\frac{1}{\Gn_{\rm sc}}
+\Gd e^{i\phi}\Gn_{\rm sc}\left(\frac{r_{\rm c}}{r_{\rm s}}\right)^{2n}\right]\left[\left(\frac{r_{\rm c}}{r_{\rm s}}\right)^{2n}+\Gd e^{i\phi}\right]^{-1}
\frac{1}{n}\left[\left(\frac{z\overline{z_0}}{r_{\rm s}^2}\right)^{-n}+\left(\frac{\overline{z}z_0}{r_{\rm s}^2}\right)^{-n}\right] \nonum & ~&
\eeqa{0.11}
where, in accordance with the definitions in \citeAPY{Milton:2005:PSQ} we have introduced the real parameters $\phi$ and
$\Gd$ (not to be confused with the delta function) and the complex parameter $\Gn_{\rm sc}$ defined via
\beq \Gd e^{i\phi}=\frac{(\Gve_{\rm s}+\Gve_{\rm c})(\Gve_{\rm m}+\Gve_{\rm s})}{(\Gve_{\rm s}-\Gve_{\rm c})(\Gve_{\rm m}-\Gve_{\rm s})},\quad
\Gn_{\rm sc}=\frac{\Gve_{\rm s}-\Gve_{\rm c}}{\Gve_{\rm s}+\Gve_{\rm c}}
\eeq{0.12}
These expressions for $V_{\rm c}$, $V_{\rm s}$ and $V_{\rm m}$ are valid both for the cases $r_{\rm s}>r_{\rm c}$ and $r_{\rm c}>r_{\rm s}$.

\begin{figure}
\vspace{1in}
\begin{minipage}[b]{3.0in}
{\resizebox{3.0in}{3.0in}
{\includegraphics{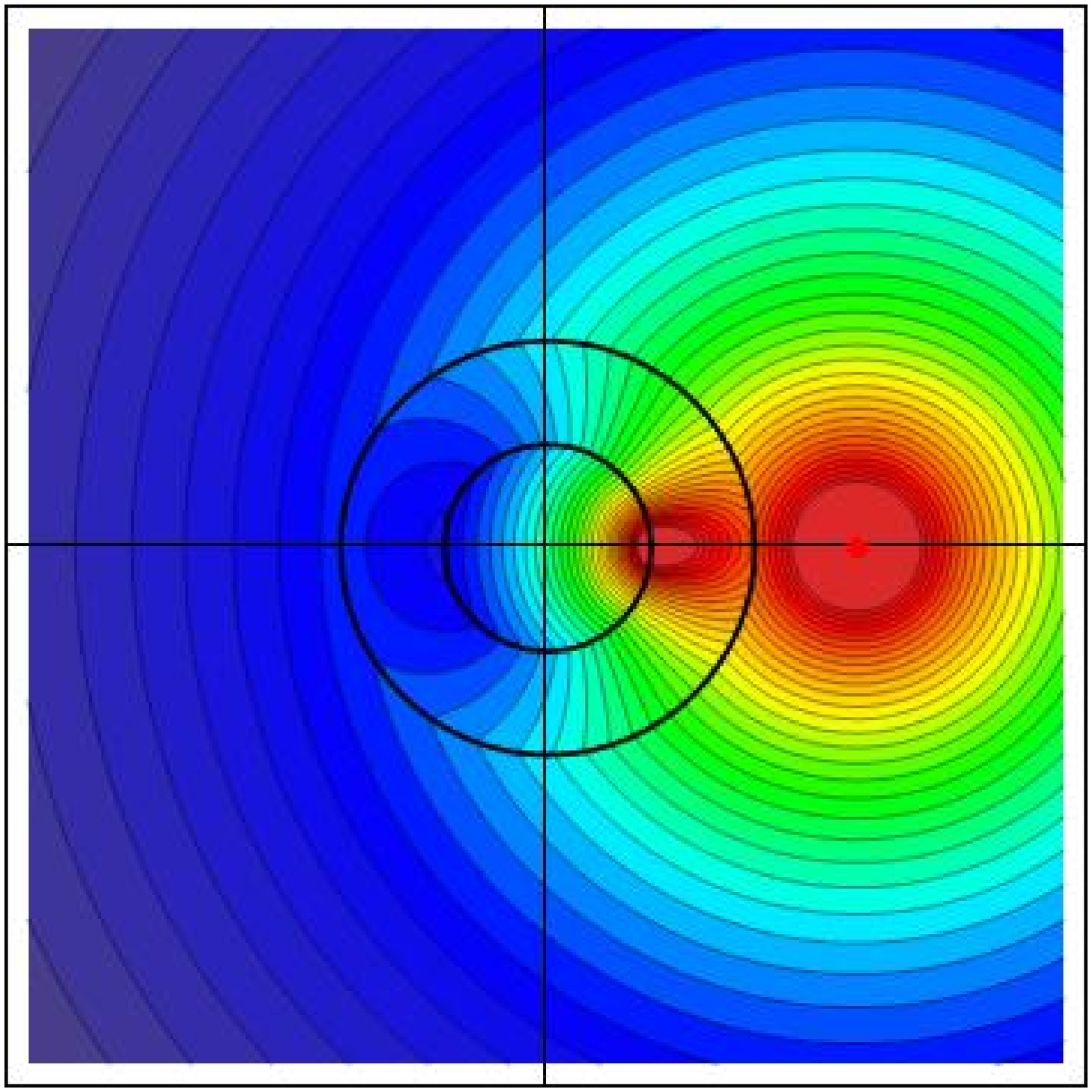}}}
\end{minipage}
~~
\begin{minipage}[b]{3.0in}
{\resizebox{3.0in}{3.0in}
{\includegraphics{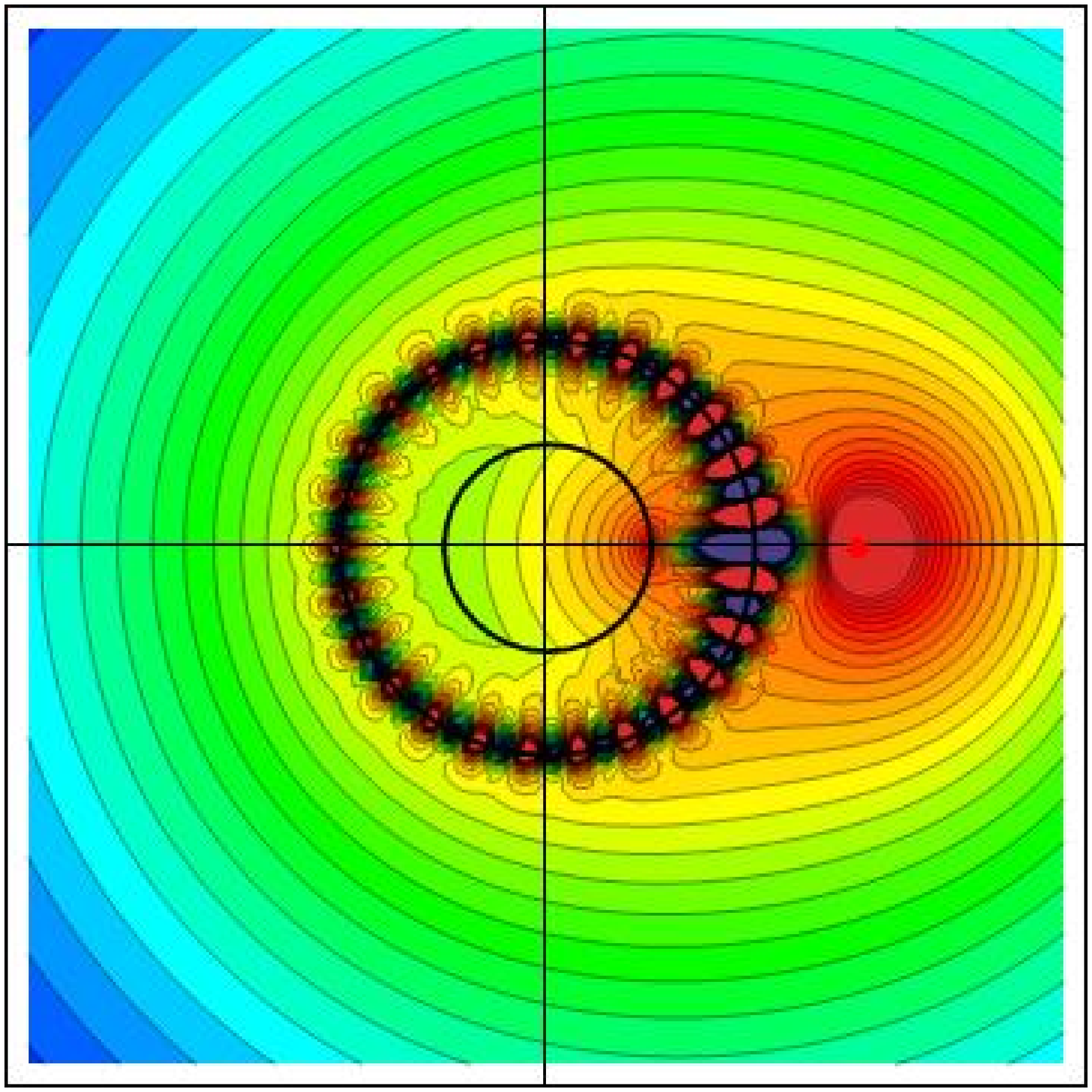}}}
\end{minipage}
\vspace{0.1in} \\
\begin{minipage}[b]{3.0in}
{\resizebox{3.0in}{0.3in}
{\includegraphics{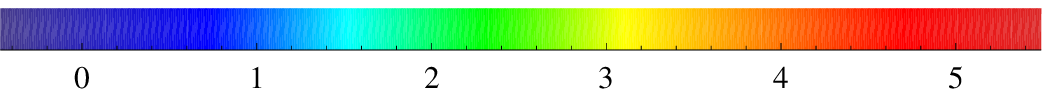}}}
\end{minipage}
~~
\begin{minipage}[b]{3.0in}
{\resizebox{3.0in}{0.3in}
{\includegraphics{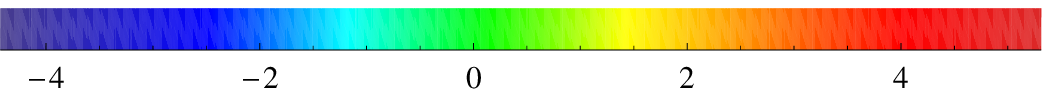}}}
\end{minipage}
\vspace{0.05in}
\caption{Numerical computations for the potential associated with a monopole
at $z_0=6$
in the unfolded geometry (unfolding parameter $a=0.7$) with $\Gve_{\rm s}=-1+ 10^{-9} i$,  (a)$\Gve_{\rm c}=\Gve_{\rm m}=1$
and (b)$\Gve_{\rm c}=5$, $\Gve_{\rm m}=1$. In both cases,
$r_{\rm c}=5.4$, $r_{\rm c}'=2$, $r_{\rm s}=r_{\rm s}'=4$.}
\labfig{1}
\end{figure}

In Fig.2 we show the potential around a monopole when mapped to the unfolded geometry. The contrast is evident between the case of
a core of dielectric constant matching that of the matrix, which is non-resonant in this example, and the case when $\Gve_{\rm c}\neq\Gve_{\rm m}$,
which exhibits anomalous resonance. Note that in the first case the coated inclusion is almost invisible: the equipotentials outside it
are nearly circular.

By letting $z_0=r_0 e^{i\Gt_0}$ and differentiating \eq{0.11} with respect to $r_0$ and with
respect to $\Gt_0$ one obtains formula for the potential associated with
a dipole at $z_0$ oriented in the
radial direction, and with one oriented in the tangential direction. The potential associated
with an arbitrarily oriented dipole is of course a linear combination of these two potentials and
is given by the formulae
\beqa V_{\rm c}& = &  (k^{(1)}+ k^{(2)})/r_0+k^{(1)}F_{\rm c}(z,z_0)+k^{(2)}F_{\rm c}(\overline{z},\overline{z_0}) \nonum
V_{\rm s} & =& (k^{(1)}+ k^{(2)})/r_0+k^{(1)}\underline{F}_{\rm out}(z,z_0)
+k^{(2)}\underline{F}_{\rm out}(\overline{z},\overline{z_0})
+ k^{(2)}\underline{F}_{\rm in}(z,z_0)+k^{(1)}\underline{F}_{\rm in}(\overline{z},\overline{z_0}) \nonum
V_{\rm m} & = & \frac{k^{(1)}}{r_0(1-z/z_0)}+\frac{k^{(2)}}{r_0(1-\overline{z}/\overline{z_0})}
+ k^{(2)}F_{\rm in}(z,z_0)+k^{(1)}F_{\rm in}(\overline{z},\overline{z_0})
\nonum &~&
\eeqa{0.13}
where
\beq k^{(1)}=(-k^{\rm e}+ik^{\rm o})/2,\quad k^{(2)}=-(k^{\rm e}+ik^{\rm o})/2
\eeq{0.14}
in which [in accordance with the definition below equation (3.5) in \citeAPY{Milton:2006:CEA}]
$k^{\rm e}$ and $k^{\rm o}$ are the (generally complex) suitably normalized amplitudes of the dipole components which have
even and odd symmetry about the line $\Gt=\Gt_0$, and
\beqa F_{\rm c}(z,z_0) &= &\frac{4\Gve_{\rm s}\Gve_{\rm m}}{r_0(\Gve_{\rm s}-\Gve_{\rm c})(\Gve_{\rm m}-\Gve_{\rm s})}S(\Gd,hz/z_0)\nonum
\underline{F}_{\rm out}(z,z_0)& = &\frac{2\Gve_{\rm m}}{r_0\Gn_{\rm sc}(\Gve_{\rm m}-\Gve_{\rm s})}S(\Gd,hz/z_0) \nonum
\underline{F}_{\rm in}(z,z_0)& = & \frac{2\Gve_{\rm m}}{r_0(\Gve_{\rm m}-\Gve_{\rm s})}S(\Gd,r_{\rm s}^2/(z\overline{z_0}))\nonum
F_{\rm in}(z,z_0)& = & \frac{S(\Gd, r_{\rm s}^4/(r_{\rm c}^2z\overline{z_0}))}{r_0\Gn_{\rm sc}}
+\frac{\Gd e^{i\phi} \Gn_{\rm sc} S(\Gd, r_{\rm s}^2/(z\overline{z_0}))}{r_0}
\eeqa{0.15}
in which
\beq h=\frac{r_{\rm s}^2}{r_{\rm c}^2}, \quad
S(\Gd, w)=\sum_{\ell=1}^\infty\frac{w^{\ell}}{1+\Gd e^{i\phi} h^{\ell}},
\eeq{0.16}
and the remaining functions are obtained by replacing $z$ and $z_0$ with $\overline{z}$ and $\overline{z_0}$
in \eq{0.15}.
These formulae for the potentials agree with the formulae of \citeAPY{Milton:2005:PSQ}
and (for a dipole not on the $x_1$ axis) with the formulae in the supporting online
material of \citeAPY{Nicorovici:2007:OCT} (see http://www.physics.usyd.edu.au/cudos/research/plasmon.html)
aside from the (irrelevant) additive constant of $(k^{(1)}+ k^{(2)})/r_0$.

It is interesting to see what happens to the potential in the matrix in the limit as $\Gve_{\rm s}$ approaches $\Gve_{\rm m}$.
Specifically, let us suppose that $k^{(1)}$, $k^{(2)}$, $\Gve_{\rm c}$, and $\Gve_{\rm m}$ remain fixed
with $\Gve_{\rm m}$ real and positive, and with $\Gve_{\rm c}$ possibly complex
(with non-negative imaginary part) but not real and negative,
and that $\Gve_{\rm s}$ approaches $\Gve_{\rm m}$ along a trajectory in the lower half of the
complex plane in such a way that $\Gd\to \infty$ but $\phi$ remains fixed. We set
\beq
\Gn=\frac{\Gve_{\rm m}-\Gve_{\rm c}}{\Gve_{\rm m}+\Gve_{\rm c}}
\eeq{1.0aa}

When $\Gve_{\rm s}$ is close to $\Gve_{\rm m}$ \eq{0.12} implies
\beqa \Gve_{\rm s} & \approx & [1-2e^{-i\phi}/(\Gd\Gn)]\Gve_{\rm m}\quad {\rm when}\quad \Gve_{\rm c} \ne \Gve_{\rm m}, \nonum
 & \approx & (1-2ie^{-i\phi/2}/\sqrt{\Gd})\Gve_{\rm m}\quad {\rm when}\quad \Gve_{\rm c} = \Gve_{\rm m},
\eeqa{1.0a}
and so we have
\beqa \Gd & \approx & 2\Gve_{\rm m}/(|\Gve_{\rm s}-\Gve_{\rm m}||\Gn|) \quad {\rm when}\quad \Gve_{\rm c} \ne \Gve_{\rm m}, \nonum
 & \approx & 4\Gve_{\rm m}^2/|\Gve_{\rm s}-\Gve_{\rm m}|^2 \quad      {\rm when}\quad \Gve_{\rm c} = \Gve_{\rm m}.
\eeqa{1.0b}
Thus for large $\Gd$ the trajectory approaches $\Gve_{\rm m}$ in such a way that the
argument of $\Gve_{\rm s}-\Gve_{\rm m}$ is approximately constant. Since
the imaginary part of $\Gve_{\rm s}$ is strictly negative, while the imaginary
part of $\Gn$ is negative or zero, we deduce that $\phi$ is not equal
to $\pi$ or $-\pi$ and this ensures that there are no infinite terms
in the series \eq{0.16}.

We need an approximation for $S(\Gd,w)$ in the limit where $\Gd$ is very large. From \eq{0.16} we see
that when $|w|<h$ the series expansion for $\Gd S(\Gd,w)$ converges in the
limit $\Gd\to\infty$ and as a consequence
\beq S(\Gd,w)\approx \frac{e^{-i\phi}w}{\Gd(h-w)},
\eeq{1.9aaa}
When $1>|w|>h$ the terms in the series for $S(\Gd,w)$ first increase
exponentially until $\ell$ reaches a transition region where $\ell\approx n$
in which $n$ is the largest integer such that $\Gd h^n\geq 1$ and after this
transition region the terms in the series decay exponentially. To a good approximation
(which becomes better as $\Gd\to\infty$) we have
\beq S(\Gd,w)= w^n\sum_{\ell=1}^{\infty}\frac{w^{\ell-n}}{1+\Gd e^{i\phi}h^\ell}\approx
w^n\sum_{j=-\infty}^{\infty}\frac{w^j}{1+\Gd h^n e^{i\phi}h^j},
\eeq{sapprox}
Since $\Gd h^n \to 1$ as $\Gd\to\infty$, upon solving for $n$ in terms of $h$ and $\delta$ we obtain
\beq S(\Gd,w) \approx  e^{-\log w \log\Gd/\log h}T(w)
\eeq{1.9aa}
where
\beq
T(w)=\sum_{j=-\infty}^{\infty}\frac{w^j}{1+e^{i\phi}h^j}.
\eeq{1.9ab}

Assuming $z$ is in the matrix, let us first treat the case when $\Gve_{\rm c}\ne\Gve_{\rm m}$.
Then as $\Gd\to\infty$, $\Gn_{\rm sc}$ approaches $\Gn$ and for
$h>r_{\rm s}^2/|z\overline{z_0}|$,
i.e. for $|z|>r_{\rm c}^2/r_0$, \eq{1.9aaa} implies
\beq \lim_{\Gd\to \infty} F_{\rm in}(z,z_0)=\widetilde{F}_{\rm in}(z,z_0)
\equiv \frac{\Gn r_{\rm c}^2}{r_0(z\overline{z_0}-r_{\rm c}^2)}
\eeq{0.17}
and as a consequence the potential $V_{\rm m}$ in the matrix,
with $|z|>r_{\rm c}^2/r_0$ approaches
\beqa \widetilde{V}_{\rm m} & = &
k^{(1)}\left[\frac{1}{r_0(1-z/z_0)}+\frac{\Gn r_{\rm c}^2}{r_0(\overline{z}z_0-r_{\rm c}^2)}\right]\nonum
& ~ & +k^{(2)}\left[\frac{1}{r_0(1-\overline{z}/\overline{z_0})}
+\frac{\Gn r_{\rm c}^2}{r_0(z\overline{z_0}-r_{\rm c}^2)}\right]
\eeqa{0.18}
which, as might be expected, is exactly the same potential which would be associated with
line dipole outside a solid cylinder of dielectric constant $\Gve_{\rm c}$ and radius $r_{\rm c}$.
In the unfolded geometry it appears as if shell has the effect of magnifying the core by the
factor $r_{\rm c}/r_{\rm c}'=r_{\rm c}/(r_{\rm s}-a(r_{\rm c}-r_{\rm s}))$. When the source is located with
$r_{\rm s}<r_0<r_{\rm c}^2/r_{\rm s}$ it will look like there is a ghost singularity in the
matrix positioned at $z=r_{\rm c}^2/\overline{z_0}$. When $r_{\rm s}<|z|<r_{\rm c}^2/r_0$ \eq{1.9aa} implies
$\Gd S(\Gd,r_{\rm s}^2/(z\overline{z_0}))$ scales like $\Gd^{\log(r_{\rm c}^2/(|z|r_0))/(-\log h)}$
and as a result this is a region of anomalous resonance with the potential $V_{\rm m}$ diverging
inside it, with this same scaling.

When $\Gve_{\rm c}=\Gve_{\rm m}$ the same argument shows that as, $\Gd\to\infty$, $F_{\rm in}(z,z_0)$ tends to zero
for $|z|>r_{\rm c}^2/r_0$. In fact it converges to zero in a larger region. To see this,
note that $\Gn_{\rm sc}$ scales as $1/\sqrt{\Gd}$, and as a consequence of which
$\Gd \Gn_{\rm sc}S(\Gd,r_{\rm s}^2/(z\overline{z_0}))$ scales like $\Gd^{\Gj}$
where $\Gj=\log(r_{\rm c}r_{\rm s}/r_0|z|)/(-\log h)$,
in the region $r_{\rm s}<|z|<r_{\rm c}^2/r_0$. This converges to zero for $|z|>r_\#^2/r_0$, where
$r_\#=\sqrt{r_{\rm c}r_{\rm s}}$, but diverges to infinity (with increasingly rapid spatial
oscillations) in the region $r_{\rm s}<|z|<r_\#^2/r_0$. Thus, as $\Gd\to\infty$,  the potential $V_{\rm m}$ will converge
for $|z|>r_\#^2/r_0$  to the potential associated with a line dipole in free space, while
diverging to infinity in the anomalously resonant region $r_{\rm s}<|z|<r_\#^2/r_0$.

It is also interesting to consider the limit as $\Gve_{\rm s}$ approaches $-\Gve_{\rm m}$ in the folded geometry. The results of
\citeAPY{Nicorovici:1994:ODP} apply directly to this case, and show that the coated cylinder in the folded geometry is equivalent to a solid cylinder
of dielectric constant $\Gve_{\rm c}$ of radius $r_{\rm s}^2/r_{\rm c}$, which is less than $r_{\rm s}$. In particular, in the unfolded geometry,
the inclusion will be invisible when $\Gve_{\rm c}=\Gve_{\rm m}$: presumably such an object acts as a lens to shrink the apparent size of any
object inside it. One can check that anomalous resonance and cloaking do not occur
for sources outside the inclusion in this circumstance.
\section{Cloaking of a single polarizable line dipole}
\setcounter{equation}{0}
First we present an example which shows that a polarizable line
with polarizability $\Ga$  can be cloaked when immersed in a TM field
surrounding a folded coated cylinder with core radius $r_{\rm c}$ and shell radius $r_{\rm s}<r_{\rm c}$
and with cylinder axis $x_1=x_2=0$.
The polarizable line is placed along $x_1=r_0$ and $x_2=0$, where $r_0>r_{\rm s}$.
Suppose $(E_1(x_1,x_2),E_2(x_1,x_2),0)$ is the field with the
polarizable line absent (but with the coated cylinder present) due to
fixed sources not varying in the $x_3$ direction lying outside the radius
$r_{\rm c}$ when $\Gve_{\rm c}\ne \Gve_{\rm m}$, and the radius $r_\#\equiv\sqrt{r_{\rm s} r_{\rm c}}$
when $\Gve_{\rm c}=\Gve_{\rm m}$. We assume these sources are not perturbed when
the polarizable line is introduced.

Again, let us suppose that $\Gve_{\rm c}$ and $\Gve_{\rm m}$ remain fixed and that $\Gve_{\rm s}$ approaches $\Gve_{\rm m}$
along a trajectory in the lower half of the
complex plane in such a way that $\Gd\to \infty$ but $\phi$ remains fixed.
Let us drop the $E_3$ field component of the electric field since it is zero for TM fields.
The field $(E_1^0,E_2^0)$ acting on the polarizable line has two components:
\beq (E_1^0, E_2^0)=(E_1+E_1^{\rm r}, E_2+E_2^{\rm r}),
\eeq{1.1}
where
\beqa &~& E_1 \equiv  E_1(r_0,0),\quad  E_2\equiv E_2(r_0,0),
\quad  E_1^{\rm r}\equiv E_1^{\rm r}(r_0,0), \quad  E_2^{\rm r}\equiv E_2^{\rm r}(r_0,0), \nonum
 &~& (E_1^{\rm r}(x,y),E_2^{\rm r}(x,y))  =  (-\Md V_{\rm in}(x_1,x_2)/\Md x_1,-\Md V_{\rm in}(x_1,x_2)/\Md x_2),
\eeqa{1.2}
and $V_{\rm in}(x_1,x_2)$ is the (possibly resonant) response potential in the matrix
generated by the coated
cylinder responding to the polarizable line itself (not including
the field generated by the coated cylinder responding to the other fixed sources).
From  \eq{0.13}, \eq{0.14} and \eq{0.15}, or alternatively from
(2.5), (3.9) and (3.10) of \citeAPY{Milton:2005:PSQ}, we have
\beq V_{\rm in}(x_1,x_2)=  [f_{\rm in}^{\rm e}(z)+f_{\rm in}^{\rm e}(\bar{z})]/2+[f_{\rm in}^{\rm o}(z)-f_{\rm in}^{\rm o}(\bar{z})]/(2i),
\eeq{1.3}
where $z=x_1+i x_2$ and for p$=$e,o
\beq f_{{\rm in}}^{\rm p}(z)=-qk^{\rm p}F_{\rm in}(z,r_0)
=-\frac{qk^{\rm p}S(\Gd,r_*^2/(r_0z))}{r_0\Gn_{\rm sc}}-\frac{qk^{\rm p}\Gd e^{i\phi}\Gn_{\rm sc}S(\Gd,r_{\rm s}^2/(r_0z))}{r_0},
\eeq{1.3a}
in which $k^{\rm e}$ and $k^{\rm o}$ are the (suitably normalized) dipole moments of the
polarizable line ($k^{\rm e}$ gives the amplitude of the dipole component which has
even symmetry about the $x_1$-axis while $k^{\rm o}$ gives the amplitude of the dipole component which has
odd symmetry about the $x_1$-axis ) and in which
$q=1$ for p=e and $q=-1$ for p=o.
Differentiating \eq{1.3} gives
\beqa  E_1^{\rm r}(x_1,x_2) & = & -[{f_ {\rm in}^{\rm e}}'(z)+{f_{\rm in}^{\rm e}}'(\bar{z})]/2
                        -[{f_{\rm in}^{\rm o}}'(z)-{f_{\rm in}^{\rm o}}'(\bar{z})]/(2i), \nonum
 E_2^{\rm r}(x,y)& = & -i[{f_{\rm in}^{\rm e}}'(z)-{f_{\rm in}^{\rm e}}'(\bar{z})]/2
                 -[{f_{\rm in}^{\rm o}}'(z)+{f_{\rm in}^{\rm o}}'(\bar{z})]/2,
\eeqa{1.5a}
where
\beq
{f_{\rm in}^{\rm p}}'(z)\equiv df_{{\rm in}}^{\rm p}(z)/dz
=\frac{qk^{\rm p} r_*^2 S'(\Gd,r_*^2/(r_0z))}{r_0^2 z^2\Gn_{\rm sc}}
+\frac{qk^{\rm p} r_{\rm s}^2\Gd e^{i\phi}\Gn_{\rm sc}S'(\Gd,r_{\rm s}^2/(r_0z))}{r_0^2z^2},
\eeq{1.5aa}
in which
\beq S'(\Gd,w)\equiv \frac{dS(\Gd,w)}{dw} = \sum_{\ell=1}^{\infty}\frac{\ell w^{\ell-1}}{1+\Gd e^{i\phi}h^\ell}.
\eeq{1.5ab}
These expressions simplify if $z$ is real since then ${f_{\rm in}^{\rm p}}'(z)-{f_{\rm in}^{\rm p}}'(\bar{z})=0$
and $(E_1^{\rm r},E_2^{\rm r})=(-{f_{\rm in}^{\rm e}}'(z),-{f_{\rm in}^{\rm o}}'(z))$.
In particular with $z=r_0$ we obtain
\beq \pmatrix{E_1^{\rm r} \cr E_2^{\rm r}}=c(\Gd)\pmatrix{k^{\rm e} \cr -k^{\rm o}},
\eeq{1.6}
where
\beq c(\Gd)
=-\frac{r_*^2 S'(\Gd,r_*^2/r_0^2)}{r_0^4\Gn_{\rm sc}}
-\frac{r_{\rm s}^2\Gd e^{i\phi}\Gn_{\rm sc}S'(\Gd,r_{\rm s}^2/r_0^2)}{r_0^4}.
\eeq{1.6aa}
We will see that $|c(\Gd)|$ can diverge to infinity as $\Gd\to \infty$, and
that when this happens the polarizable line becomes cloaked.

Now if $\Ga$ denotes the polarizability of the line, then we have
\beq \pmatrix{k^{\rm e} \cr -k^{\rm o}} = \Ga\pmatrix{E_1^0 \cr E_2^0}.
\eeq{1.7}
This implies
\beq \pmatrix{k^{\rm e} \cr -k^{\rm o}}=\Ga\pmatrix{E_1 \cr E_2}+\Ga c(\Gd)\pmatrix{k^{\rm e} \cr -k^{\rm o}},
\eeq{1.7a}
which when solved for the dipole moment $(k^{\rm e},-k^{\rm o})$ gives
\beq \pmatrix{k^{\rm e} \cr -k^{\rm o}} = \Ga_*\pmatrix{E_1 \cr E_2},
\eeq{1.8}
where
\beq \Ga_*=[\Ga^{-1}-c(\Gd)]^{-1},
\eeq{1.9}
is the ``effective polarizability''. So far no approximation has been made.

Notice that when $|c(\Gd)|$ is very large then $\Ga_* \approx -1/c(\Gd)$.
So in this limit the effective polarizability has a very weak dependence on $\Ga$.

To obtain an asymptotic formula for $c(\Gd)$ when $\Gd$ is very large we
use the asymptotic formula \eq{1.9aaa} and \eq{1.9aa}. Differentiating
these gives
\beq S'(\Gd,w)\approx \frac{e^{-i\phi}h}{\Gd(h-w)^2}~~{\rm for}~|w|<h
\eeq{1.9aaaN}
for $|w|<h$ while when $1>|w|>h$
\beqa
S'(\Gd,w)& \approx & -[\log\Gd/(w\log h)]e^{-\log w \log\Gd/\log h}T(w)+e^{-\log w \log\Gd/\log h}T'(w) \nonum
& \approx & -[\log\Gd/(w\log h)]e^{-\log w \log\Gd/\log h}T(w),
\eeqa{1.9aaN}
where $T'(w)=dT(w)/dw$
and in making the last approximation in \eq{1.9aaN} we have assumed that
$|\log\Gd|$ is very large. Let us first
treat the case where $\Gve_{\rm c}$ is fixed and not equal to $\Gve_{\rm m}$ and  $r_0<r_{\rm c}$.
Then we have $\Gn_{\rm sc}\approx \Gn$ and
substituting these approximations in \eq{1.3a} and \eq{1.6aa}
and keeping only the terms which are dominant
because $\Gd$ is very large gives, for $r_{\rm c}^2/r_0>|z|>r_{\rm s}$,
\beq f_{\rm in}^{\rm p}(z)\approx -qk^{\rm p}\Gn e^{i\phi}e^{[\log z-\log(r_{\rm c}^2/r_0)]\log\Gd/\log h}r_0^{-1}T(r_{\rm s}^2/(r_0z)),
\eeq{1.9ac}
which implies
\beq {f_{\rm in}^{\rm p}}'(z)\approx\frac{-qk^{\rm p}\Gn e^{i\phi}\log\Gd}
    {zr_0\log h}e^{-\log(r_{\rm c}^2/(zr_0)\log\Gd/\log h}T(r_{\rm s}^2/(r_0z)),
\eeq{1.9ad}
and
\beq c(\Gd)\approx \frac{\Gn e^{i\phi}\log\Gd}{r_0^2\log h}e^{-2\log(r_{\rm c}/r_0)\log\Gd/\log h}T(r_{\rm s}^2/r_0^2).
\eeq{1.9ae}
We see that $|c(\Gd)|\to\infty$ as $\Gd\to \infty$ when $r_0<r_{\rm c}$. Thus
for a polarizable line dipole inside the radius $r_{\rm c}$
the ``effective polarizability''  approaches zero in the limit $\Gd\to\infty$.
When $\Gd$ is very large from \eq{1.8} and \eq{1.9} we have
\beq k^{\rm e}\approx -E_1/c(\Gd),\quad k^{\rm o}\approx E_2/c(\Gd).
\eeq{1.9b}
Thus for $z$ in the annulus $r_c^2/r_0>|z|>r_s$
the potential associated with the polarizable line
has, from \eq{1.9ac},
\beq
f_{\rm in}^{\rm e}
        \approx  E_1
\Gd^{\log(z/r_0)/\log h}r_0T(r_{\rm s}^2/(r_0z))\log h/(T(r_{\rm s}^2/r_0^2)\log\Gd).
\eeq{1.9c}
Similarly in this annulus we have
\beq
f_{\rm in}^{\rm o} \approx  E_2
\Gd^{\log(z/r_0)/\log h}r_0T(r_{\rm s}^2/(r_0z))\log h/(T(r_{\rm s}^2/r_0^2)\log\Gd).
\eeq{1.10}
For $z$ outside the radius $r_{\rm c}^2/r_0$ the potential due to the polarizable
line dipole is approximately given by \eq{0.18} and converges to zero because
$k^{(1)}$ and $k^{(2)}$ vanish as $\Gd\to \infty$. We avoid the technical
question of what happens when $|z|=r_{\rm c}^2/r_0$  but presumably the potential
also converges to zero there.

Thus as $\Gd\to \infty$ the potential in the matrix due to the polarizable
line dipole converges to zero in the region $r>r_0$
but diverges to infinity with increasingly rapid angular oscillations for $r_{\rm s}\leq r<r_0$.
(This is to be contrasted with the potential in the matrix associated with a line dipole
having fixed $k^{\rm e}$ and $k^{\rm o}$,
which as can be seen from \eq{1.9ac} diverges to infinity in the much larger region
$r_{\rm s}\leq r <r_{\rm c}^2/r_0$.) A simple calculation shows that in the shell the potential associated with the
polarizable line similarly converges to zero for $r>r_0$ but diverges to infinity for $r_{\rm s}<r<r_0$,
while in the core the potential associated with the polarizable line converges to zero everywhere.

It is instructive to see what happens to
the local field $(E_1^0,E_2^0)$ acting on the polarizable line as $\Gd\to \infty$.
From \eq{1.1}, \eq{1.6}, \eq{1.8} and \eq{1.9} we see that
\beq E_1^0=E_1+c(\Gd)k^{\rm e}=E_1+\frac{c(\Gd)E_1}{\Ga^{-1}-c(\Gd)}=\frac{E_1}{1-\Ga c(\Gd)}
\eeq{1.10a}
goes to zero as $\Gd\to \infty$, and similarly so too does $E_2^0$.  This explains why the ``effective polarizability''
vanishes as $\Gd\to \infty$: the effect of the resonant field is to cancel the field $(E_1^0, E_2^0)$
acting on the polarizable line.

Suppose the source outside is a line dipole with a fixed source term $(k^{\rm e}_1,k^{\rm o}_1)=(k^{\rm e}_1,0)$
located at the point $(r_1,0)$, where $r_1>r_{\rm c}>r_0>r_{\rm s}$. When $r_1$ is chosen with
$r_{\rm c}^2/r_0>r_1>r_{\rm c}$ the polarizable line will be located within the resonant
region generated by the line source outside. One might at first think that a polarizable
line placed within the resonant region would have a huge response because of the
enormous fields there. However, we will see that the opposite is true: the
dipole moment of the polarizable line still goes to zero as $\Gd\to \infty$.
From \eq{1.6}, \eq{1.5a} and \eq{1.9ad}, with $r_0$ replaced by $r_1$, the field at the point $(r_0,0)$
when the polarizable line is absent will be
\beq E_1 = c_1(\Gd)k^{\rm e}_1, \quad E_2=0,
\eeq{1.11}
where
\beq c_1(\Gd)\approx \frac{\Gn e^{i\phi}\log\Gd}{r_0r_1\log h}e^{-\log(r_{\rm c}^2/(r_0r_1))\log\Gd/\log h}T(r_{\rm s}^2/(r_0r_1)).
\eeq{1.12}
This and \eq{1.9b} implies the polarizable line has a dipole moment
\beq
k^{\rm e}\approx -E_1/c(\Gd)
\approx - c_1(\Gd)k^{\rm e}_1/c(\Gd)
\approx -\frac{r_0T(r_{\rm s}^2/(r_0r_1))}{r_1T(r_{\rm s}^2/r_0^2)}\Gd^{\log(r_1/r_0)/\log h}k^{\rm e}_1.
\eeq{1.13}
So $k^{\rm e}$ scales as $\Gd^{\log(r_1/r_0)/\log h}$ which goes to zero (since $h<1$)
as $\Gd\to \infty$ but fairly slowly when $r_1$ and $r_0$ are almost equal, i.e. both close to $r_{\rm c}$.

If the source is outside the critical radius
$r_{\rm crit}=r_{\rm c}^2/r_{\rm s}$ then there are no resonant regions associated with
it and $k^{\rm e}$ will scale like $1/c(\Gd)$, i.e. as
$\Gd^{2\log(r_{\rm c}/r_0)/\log h}/\log\Gd$ which goes to zero
at a faster rate as $\Gd\to \infty$, but still slowly when $r_0$ is close
to $r_{\rm c}$. On the other hand when $r_0$ is close to $r_{\rm s}$ we
have $r_{\rm c}/r_0\approx 1/\sqrt{h}$ and this latter scaling is approximately
$\Gd^{-1}/\log\Gd\sim -\Gve_{\rm s}''/\log\Gve_{\rm s}''$, where
$\Gve_{\rm s}''$ is the imaginary part of $\Gve_{\rm s}$, which is quite fast.

The asymptotic analysis is basically similar when $\Gve_{\rm c}=\Gve_{\rm m}$ and $r_0<r_\#\equiv\sqrt{r_{\rm s}r_{\rm c}}$.
Then $\Gn_{\rm sc}\approx -ie^{-i\phi/2}/\sqrt{\Gd}$ and from \eq{1.3a}, \eq{1.6aa}, \eq{1.9aa}, and \eq{1.9aaN}
we have for $r_{\rm c}^2/r_0>|z|>r_{\rm s}$ that
\beq f_{\rm in}^{\rm p}(z)\approx i qk^{\rm p} e^{i\phi/2} e^{[\log z-\log(r_{\rm c} r_{\rm s}/r_0)]\log\Gd/\log h}r_0^{-1}T(r_{\rm s}^2/(r_0z)),
\eeq{1.14}
and
\beq c(\Gd)\approx \frac{-i e^{i\phi/2}\log\Gd}{r_0^2\log h}e^{-\log(r_{\rm c}r_{\rm s}/r_0^2)\log\Gd/\log h}T(r_{\rm s}^2/r_0^2).
\eeq{1.15}
When all the sources lie outside the critical radius $r_{\rm c}$ so they do not generate any
resonant regions in the absence of the polarizable line, both $k^{\rm e}$ and $k^{\rm o}$ will
scale as $1/c(\Gd)$, i.e. as $\Gd^{\log(r_{\rm c}r_{\rm s}/r_0^2)/\log h}/\log\Gd$, as $\Gd\to \infty$.
When $r_0$ is close to $r_{\rm s}$ we have $r_{\rm c}r_{\rm s}/r_0^2 \approx 1/\sqrt{h}$ and this latter scaling is approximately
$1/(\sqrt{\Gd}\log\Gd)\sim-\Gve_{\rm s}''/\log\Gve_{\rm s}''$ which is the same as when $\Gve_{\rm c}\ne\Gve_{\rm m}$.
By substituting \eq{1.9b} in \eq{1.14} we obtain
\beqa  f_{\rm in}^{\rm e}(z) &\approx & -E_1
ie^{i\phi/2} e^{[\log z-\log(r_{\rm c}r_{\rm s}/r_0)]\log\Gd/\log h}r_0^{-1}T(r_{\rm s}^2/(r_0z))/c(\Gd) \nonum
  &\approx &  E_1
\Gd^{\log(z/r_0)/\log h}r_0T(r_{\rm s}^2/(r_0z))\log h/(T(r_{\rm s}^2/r_0^2)\log\Gd), \nonum
&~&
\eeqa{1.16}
which coincides with \eq{1.9c}.
Likewise \eq{1.10} still holds. By similar arguments applied to $V_{\rm c}$ and $V_{\rm s}$
it follows that
as $\Gd\to \infty$ the potential $V$
diverges with increasingly rapid oscillations in the core in the region $r_{\rm c}>r>r_{\rm c}r_{\rm s}/r_0$,
in the shell in the two regions $r_{\rm s}<r<r_0$ and $r_{\rm c}>r>r_{\rm c}r_{\rm s}/r_0$, and in the matrix in the
region $r_{\rm s}<r<r_0$. Outside these regions it converges to the potential  generated by the fixed sources.

It is possible to get any cloaking radius between $r_{\rm s}$ and $r_{\rm c}$ if we let $\Gve_{\rm c}$ depend on $\Gd$,
so that $\Gve_{\rm s}-\Gve_{\rm c}$ scales as $\Gd^{-\Gb}$ and $\Gve_{\rm m}-\Gve_{\rm s}$ scales as $\Gd^{-1+\Gb}$, where
$\Gb$ is a fixed constant between 0 and 1. Then $\Gn_{\rm sc}$ will scale as $\Gd^{-\Gb}$ and $c(\Gd)$
will scale as $\Gd^\Gj\log\Gd$ with $\Gj=\log(r_{\rm c}^{2-2\Gb}r_{\rm s}^{2\Gb}/r_0^2)/(-\log h)$ and so the cloaking
radius will be $r_{\rm c}^{1-\Gb}r_{\rm s}^\Gb$. Since [based on the results of \citeAPY{Milton:2005:PSQ}
and \citeAPY{Bruno:2007:SCS}] dielectric
bodies located in the cloaking region are not perfectly imaged, it is not sufficient that $\Gve_{\rm c}$,
$\Gve_{\rm s}$, and $\Gve_{\rm m}$ be arbitrarily close to each other to ensure perfect imaging of a dielectric
body which lies inside the radius $r_{\rm c}$. Similarly, for the standard
cylindrical quasistatic superlens,
it is not sufficient that $\Gve_{\rm c}$,
$-\Gve_{\rm s}$, and $\Gve_{\rm m}$ be arbitrarily close to each other to ensure perfect
quasistatic imaging of a dielectric
body which lies inside the radius $r_*=r_{\rm s}^2/r_{\rm c}$. Also
a slab lens of thickness $d$ and permittivity $\Gve_s$ separating two media
with permittivities $\Gve_m$ and $\Gve_c$ will not necessarily provide a
good quasistatic image of a dielectric body which lies within
a distance $d$ of the slab, even when  $\Gve_{\rm c}$,
$-\Gve_{\rm s}$, and $\Gve_{\rm m}$ are arbitrarily close to each other

\section{A proof of cloaking for an arbitrary number of polarizable line dipoles}
\setcounter{equation}{0}
The concept of ``effective polarizability'' does not have much
use when two or more polarizable lines are positioned in the cloaking
region since each polarizable line will
also interact with the resonant regions generated by the other
polarizable lines and if the polarizable lines are not all on a
plane containing the coated cylinder axis then these interactions
will oscillate as $\Gd\to \infty$. However we will see here that nevertheless
the dipole moment of each polarizable line in the cloaking region must go to zero as $\Gd\to \infty$
and in such a way that no resonant field extends outside the cloaking region. This
is not too surprising. Based on the results for a single dipole line
we expect that a resonant field extending outside the cloaking region
would cost infinite energy, and the only way to avoid this is
for the dipole moment of each polarizable line in the cloaking region to go to zero as $\Gd\to \infty$.

Here we limit our attention to the cylindrical lens with the core
having approximately the same permittivity as the matrix. Also
to simplify the analysis we assume the core (but not the matrix) has some small loss.
Specifically we assume
\beq \Gve_{\rm m}=1,\quad \Gve_{\rm s}=1-i\Gk,\quad \Gve_{\rm c}=1+i\Gg\Gk, \eeq{2.0}
with $\Gk$ and $\Gg\Gk$ having positive real parts and approaching zero in such a way that
$\Gg$, which could be complex, remains fixed and $\phi$ given by \eq{0.12} also remains fixed.
In this limit \eq{0.12} implies $(\Gk+\Gg\Gk)\Gk\approx 4/(\Gd e^{i\phi})$
and since $\Gk$ and $\Gg\Gk$ have positive real parts we deduce that
$\phi$ is not equal to $\pi$ or $-\pi$. Solving for $\Gk$ we see that
\beq \Gk\approx 2e^{-i\phi/2}/\sqrt{\Gd(1+\Gg)}
\eeq{2.0b}
The potential in the core due to a single dipole in the matrix at $z_0$ is  given
by \eq{0.13} and \eq{0.15}.

If there are $m$ dipoles at
$z_1$,$z_2$,...,$z_m$ (where $z_i\ne z_j$ for all $i\ne j$)
all in the matrix then, by the superposition principle, the
potential in the core is
\beq  V_{\rm c}=\sum_{\ell=0}^{\infty}(A_{\ell}^{(c)} z^\ell+B_{\ell}^{(c)} \bar{z}^\ell),
\eeq{2.7}
where for $\ell\ne 0$
\beq A_{\ell}^{(c)} =  \frac{h^{\ell}\Gd\psi(\Gd)}{1+\Gd e^{i\phi}h^\ell}
\sum_{j=1}^{m}(k_j^{(1)}/r_j)(1/z_j)^{\ell},\quad
 B_{\ell}^{(c)} =  \frac{h^{\ell}\Gd\psi(\Gd)}{1+\Gd e^{i\phi}h^\ell}\sum_{j=1}^{m}(k_j^{(2)}/r_j)(1/\bar{z}_j)^{\ell},
\eeq{2.8}
in which $r_j=|z_j|$ and
\beq \psi(\Gd)\equiv\frac{4\Gve_{\rm s}}{\Gd (\Gve_{\rm s}-\Gve_{\rm c})(1-\Gve_{\rm s})}
\eeq{2.3}
depends on $\Gd$ through the dependence of $\Gve_{\rm s}$ and $\Gve_{\rm c}$ on $\Gd$ but tends to
$e^{i\phi}$ as $\Gd\to \infty$.

Let us suppose the dipoles positioned in the matrix at $z_1$,$z_2$,...,$z_g$
with $1\leq g\leq m$ are in the cloaking region,
while the remainder of the dipoles are outside the cloaking region, i.e.
\beq |z_j| \leq  r_\# ~~{\rm for~all}~~j\leq g, \quad\quad
       |z_j| >  r_\# ~~{\rm for~all}~~j>g,
\eeq{2.9aa}
where we allow for the special case where some
of the dipoles have $|z_j|=r_\# $: as we will see, these are also cloaked.
We do not specify how the set of dipole
moments $\{k_1,k_2,...,k_m\}$ depends on $\Gd$ except that:
\begin{itemize}
\item
We assume that each
dipole outside the cloaking region has moments which converge to fixed limits
as $\Gd\to \infty$
\beq \lim_{\Gd\to \infty}(k_j^{(1)}(\Gd), k_j^{(2)}(\Gd))=(k_{j0}^{(1)}, k_{j0}^{(2)}) \quad {\rm for~all}~~j>g.
\eeq{2.9a}
The dipole moments $k_j^{(1)}(\Gd)$ and $k_j^{(2)}(\Gd)$ inside or outside the cloaking region
are assumed to depend linearly on the field acting upon them, since non-linearities
would generate higher order frequency harmonics.
Some of them could be energy sinks, although at least one of them should be an energy source.

\item
We assume that in the unfolded geometry the energy absorbed per unit time per unit length of the coated cylinder remains
bounded as $\Gd\to \infty$,
as, for example, must be the case if the line sources only supply a finite amount
of energy per unit time per unit length. We let $W_{\rm max}$ be the maximum amount of energy
available per unit time per unit length. It is supposed that the quasistatic limit
is being taken not by letting the frequency $\Go$ tend to zero, but instead
by fixing the frequency $\Go$ and reducing the spatial size of the
system and using a coordinate system which is appropriately rescaled.
\end{itemize}

We need to show that, because the energy absorption in the core remains bounded,
the dipole moments in the cloaking region go to zero as $\Gd\to \infty$
and the resonant field does not extend outside the cloaking
region, $r\leq r_\#$. This
is certainly true when only one polarizable line is present but as cancellation
effects can occur (the energy absorption associated with two line dipoles can
be less than the absorption associated with either line dipole acting
separately) a proof is needed.

To do this we bound $k^{(1)}_{i}$ and $k^{(2)}_i$ for any given $i\leq g$ using the fact that the
energy loss within the lens is bounded by $W_{\max}$.
If $W_{\rm c}=W_{\rm c}(\Gd)$ represents the energy dissipated in the core in the unfolded geometry,
then we have the inequality
\beqa W_{\rm c} & = & (\Go/2)\int_0^{r'_{\rm c}}r'dr'\int_{0}^{2\pi}d\Gt'\thinspace \BE'(\Bx')\cdot
{\rm Imag}(\BGve')\overline{\BE'(\Bx')} \nonum
& = & (\Go/2)\Gve_{\rm c}''\int_0^{r_{\rm c}}rdr\int_{0}^{2\pi}d\Gt\thinspace \BE(\Bx)\cdot\overline{\BE(\Bx)}
\geq (\Go/2)\Gve_{\rm c}''\int_0^{r_{\rm c}}rdr\int_{0}^{2\pi}d\Gt\thinspace E_1(z)\overline{E_1(z)},
\eeqa{2.9b}
in which ${\rm Imag}$ denotes the imaginary part,
$\Gve_{\rm c}''={\rm Imag}(\Gve_{\rm c})$ and
$E_1(z)$ is the $x_1$ component of the electric field in the core in the folded geometry given by
\beq E_1(z)=-\frac{\Md V_{\rm c}}{\Md x_1}
            = -\sum_{\ell=1}^{\infty}\ell r^{\ell-1}(A_{\ell}^{(c)}e^{i(\ell-1)\Gt}
                +B_{\ell}^{(c)}e^{-i(\ell-1)\Gt}).
\eeq{2.10}
where the derivative $\Md V_{\rm c}/\Md x_1$ is calculated by
substituting $z=x_1+i x_2$ in \eq{2.7}.

Substituting this expression for the electric field back in \eq{2.9b}
and using the orthogonality properties of Fourier modes we then have
\beqa 2W_{\rm c}/\Go & \geq & 2\pi\Gve_{\rm c}''\int_0^{r_{\rm c}}dr[(A_1^{(c)}+B_1^{(c)})\overline{(A_1^{(c)}+B_1^{(c)})}r
+\sum_{\ell=2}^{\infty}\ell^2r^{2\ell-1}(A_{\ell}^{(c)}\overline{A_{\ell}^{(c)}}
+B_{\ell}^{(c)}\overline{B_{\ell}^{(c)}})]\nonum
& \geq & \pi\Gve_{\rm c}'' r_{\rm c}^2(A_1^{(c)}+B_1^{(c)})\overline{(A_1^{(c)}+B_1^{(c)})}
 +\pi\Gve_{\rm c}''\sum_{\ell=2}^{\infty}\ell r_{\rm c}^{2\ell}(A_{\ell}^{(c)}\overline{A_{\ell}^{(c)}}
+B_{\ell}^{(c)}\overline{B_{\ell}^{(c)}})\nonum
& \geq & \pi\Gve_{\rm c}''\sum_{\ell=n-m+1}^{n}\ell r_{\rm c}^{2\ell}(A_{\ell}^{(c)}\overline{A_{\ell}^{(c)}}
+B_{\ell}^{(c)}\overline{B_{\ell}^{(c)}})
=\pi\Gve_{\rm c}''\sum_{k=0}^{m-1}b_{k}(U_k\overline{U_k}+V_k\overline{V_k}),
\eeqa{2.11}
where the last identity is obtained using \eq{2.8} with the definitions
\beqa b_{k}& \equiv &(n-k)(r_*/r_{i})^{2n-2k}\Gd^2|{\psi(\Gd)}/(1+\Gd h^{n-k}e^{i\phi})|^2, \nonum
U_k& \equiv & \sum_{j=1}^{m}u_j(r_{i}/z_j)^{-k},\quad u_j\equiv(r_{i}/z_j)^n k_j^{(1)}/r_j, \nonum
V_k& \equiv & \sum_{j=1}^{m}v_j(r_{i}/\bar{z}_j)^{-k},
\quad v_j\equiv(r_{i}/\bar{z}_j)^n k_j^{(2)}/r_j,
\eeqa{2.12}
in which $k=n-\ell$, $n\geq m+1$ remains to be chosen, and $i\leq g$. From \eq{2.12} it follows that
$\BU=\BM\Bu$ and $\BV=\BM\Bv$,
where $\BM$ is the Vandermonde matrix
\beq \BM=\pmatrix{1 & 1 & 1 & \ldots  & 1 \cr
                  z_1/r_{i} & z_2/r_{i} & z_3/r_{i} &\ldots & z_m/r_{i} \cr
                  (z_1/r_{i})^2 & (z_2/r_{i})^2 &  (z_3/r_{i})^2 &\ldots & (z_m/r_{i})^2 \cr
                  \vdots & \vdots & \vdots & \ddots & \vdots \cr
                  (z_1/r_{i})^{m-1} & (z_2/r_{i})^{m-1} &  (z_3/r_{i})^{m-1} &\ldots & (z_m/r_{i})^{m-1}}.
\eeq{2.16}
From the well known formula for the determinant of a Vandermonde matrix it follows
that $\BM$ is non-singular. Therefore there exists a constant $c_i>0$
(which is the reciprocal of the norm of $\BM^{-1}$ and which only
depends on $i$, $m$ and the $z_j$) such that $|\BU|\geq c_i |\Bu|$ and $|\BV|\geq c_i |\Bv|$,
implying
\beq |\BU|^2+|\BV|^2\geq c_i^2(|\Bu|^2+|\Bv|^2) \geq
c_i^2(|u_i|^2+|v_i|^2)=c_i^2(|k_i^{(1)}|^2+|k_i^{(2)}|^2)/r_i^2
\eeq{2.17}
Next we need to select $n$ and find a lower bound on $b_{k}$ which is independent of $k$.
Let $s=-\log\Gd/\log h$ (so $\Gd h^s=1$) and take $n$ as the largest integer smaller than or equal to $s$
so $n+1\geq s\geq n$. Then
since $r_*<r_{\rm s}<r_{i}$ and $r_*=r_\# h^{3/4}$ we have
\beq (r_*/r_{i})^{2n}\Gd^2\geq (r_*/r_{i})^{2s}\Gd^2
=\Gd^2\Gd^{-2\log(r_*/r_{i})/\log h}=\Gd^{1/2}\Gd^{-2\log(r_\#/r_{i})/\log h}.
\eeq{2.18}
Also the following inequalities hold for $m-1\geq k\geq 0$
\beq 1=\Gd h^{s}\leq \Gd h^n\leq \Gd h^{n-k} \quad {\rm and} \quad
\Gd h^{n-k}\leq \Gd h^{s-k-1}\leq \Gd h^{s-m}=h^{-m}.
\eeq{2.19}
So it follows that
\beq |1+\Gd h^{n-k}e^{i\phi}|\leq a\equiv \max_{1\leq t\leq h^{-m}}|1+t e^{i\phi}|,
\eeq{2.20}
and $a$ is independent of $\Gd$. From the bounds \eq{2.18} and \eq{2.20} we deduce that
\beqa b_{k} & \geq & (s-m)(r_{i}/r_*)^{2k}\Gd^{1/2}\Gd^{-2\log(r_\#/r_{i})/\log h}|\psi(\Gd)|^2/a^2 \nonum
& \geq & -[(\log\Gd/\log h)+m]\Gd^{1/2}\Gd^{-2\log(r_\#/r_{i})/\log h}|\psi(\Gd)|^2/a^2.
\eeqa{2.21}
Combining inequalities gives
\beqa 2W_{\rm c}/\Go &\geq&
\frac{\pi\Gve_{\rm c}''|\psi(\Gd)|^2\sqrt{\Gd}}{a^2(-\log h)}(\log\Gd+m\log h)\Gd^{-2\log(r_\#/r_{i})/\log h}(|\BU|^2+|\BV|^2) \nonum
& \geq &
\frac{\pi\Gve_{\rm c}''|\psi(\Gd)|^2c_i^2\sqrt{\Gd}}{a^2r_{i}^2(-\log h)}(\log\Gd+m\log h)\Gd^{-2\log(r_\#/r_{i})/\log h}
(|k_i^{(1)}|^2+|k_i^{(2)}|^2),
\eeqa{2.22}
in which the real positive prefactor has the property that
\beq \Gr_{i}\equiv \lim_{\Gd\to \infty}\frac{\pi\Gve_{\rm c}''|\psi(\Gd)|^2c_i^2\sqrt{\Gd}}{a^2r_{i}^2(-\log h)}
=\frac{2\pi  c_i^2}{a^2r_{i}^2(-\log h)}{\rm Real}(e^{-i\phi/2}\Gg/\sqrt{1+\Gg})
\eeq{2.23}
is strictly positive, where ${\rm Real}(w)$ denotes the real part of $w$.
So there exists a $\Gd_0$ such that,
for all $\Gd>\Gd_0$ and all $i\leq g$,
\beq \frac{\pi\Gve_{\rm c}''|\psi(\Gd)|^2c_i^2\sqrt{\Gd}}{a^2r_{i}^2(-\log h)} \geq \Gr/2,\quad
{\rm where}~\Gr\equiv \min_{i\leq g}\Gr_{i}>0,
\eeq{2.24}
and such that
\beq \log\Gd+m\log h>\frac{1}{2}\log\Gd>-2\log h, \eeq{2,24a}
which, in particular, ensures that $n\geq m+1$. So we conclude that
\beq |k_i^{(1)}|^2+|k_i^{(2)}|^2
\leq 2\Gd^{\log(r_\#/r_{i})/\log h}\sqrt{2W_{\rm c}/(\Go\Gr\log\Gd)}.
\eeq{2.26}
which, since $\log(r_\#/r_{i})/\log h$ is negative,
forces the dipole amplitudes $k^{(1)}_i$ and $k^{(2)}_i$ to go to zero as $\Gd\to \infty$
(even when $r_i=r_\#$) because $W_{\rm c}=W_{\rm c}(\Gd)\leq W_{\rm max}$.

Now the superposition principle implies that the potential at any point $z$ in the matrix is
\beq V(z)=\sum_{j=1}^{m} k^{(1)}_jV_j^{(1)}(z)+ k^{(2)}_jV_j^{(2)}(z),
\eeq{2.27}
where $V_j^{(1)}(z)$ (or $V_j^{(2)}(z)$) is the potential in the matrix due to an isolated line dipole
in the matrix at the point $z_j$ with $k_j^{(1)}=1$, $k_j^{(2)}=0$
(respectively with $k_j^{(1)}=0$, $k_j^{(2)}=1$).
Now according to the analysis at the end of section 2 (which is easily extended to the case treated
here where $\Gve_{\rm c}$ depends on $\Gd$ as implied by \eq{2.0} and \eq{2.0b})
it follows that for $z$ in the matrix with $|z|>\max\{r_{\rm s}, r_\#^2/r_j\}$,
\beqa \lim_{\Gd\to \infty} V_j^{(1)}(z)=\widetilde{V}_j^{(1)}(z)\equiv \frac{1}{r_j(1-z/z_j)} \nonum
 \lim_{\Gd\to \infty} V_j^{(2)}(z)=\widetilde{V}_j^{(2)}(z)\equiv
\frac{1}{r_j(1-\overline{z}/\overline{z_j})}\eeqa{2.28}
Also, as shown in the analysis at the end of section 2, if $r_\#^2/r_j>|z|>r_{\rm s}$, then $V_j^{(1)}(z)$ and
$V_j^{(2)}(z)$ diverge as $\Gd^{\Gj}$ where $\Gj=\log(r_{\rm c}r_{\rm s}/r_j|z|)/(-\log h)$. If $z_j$ is outside the cloaking region (i.e. $j> g$) then
$r_\#^2/r_j$ will be less than $r_\#$. So using the well known fact that
\beq \lim_{\Gd\to \infty} e(\Gd)f(\Gd)=e_0f_0,\quad{\rm where}~~e_0=\lim_{\Gd\to \infty} e(\Gd), \quad
f_0=\lim_{\Gd\to \infty} f(\Gd),
\eeq{2.30}
it follows that for all $|z|>r_\#$ and all $j>g$
\beqa  \lim_{\Gd\to \infty} k^{(1)}_jV_j^{(1)}(z)=k_{j0}^{(1)}\widetilde{V}_j^{(1)}(z)\nonum
 \lim_{\Gd\to \infty} k^{(1)}_jV_j^{(2)}(z)=k_{j0}^{(1)}\widetilde{V}_j^{(2)}(z)
\eeqa{2.31}
If $z_i$ is inside the cloaking region (i.e. $i\leq g$) and $|z|>r_\#^2/r_i$ then \eq{2.28},
\eq{2.30} and the fact
that $|k^{(1)}_i|$ and $|k^{(2)}_i|$ tend to zero implies that $k^{(1)}_iV_i^{(1)}(z)$ and
$k^{(2)}_iV_i^{(2)}(z)$ will tend to zero. For
$r_\#^2/r_i>|z|>r_\#$ we have that $V_i^{(1)}(z)$ and  $V_i^{(2)}(z)$ scale as $\Gd^{\Gj}$
with $\Gj=\log(r_{\rm c}r_{\rm s}/(r_i|z|))/(-\log h)$ while from \eq{2.26} $k^{(1)}_i$
and $k^{(2)}_i$ scale at worst as $\Gd^{-t}/(\log\Gd)$
with $t=\log(r_\#/r_{i})/(-\log h)$. So their product $k^{(1)}_iV_i^{(1)}(z)$ or $k^{(2)}_iV_i^{(2)}(z)$will scale at worst as
$\Gd^{\Gj-t}/\log\Gd$ where $\Gj-t=\log(r_\#/|z|)/(-\log h)$. This {\it goes to zero} as ${\Gd\to \infty}$
when $|z|>r_\#$. By taking the limit $\Gd\to \infty$ of both sides of \eq{2.27} we conclude that
\beq  \lim_{\Gd\to \infty}V(z)
=\sum_{j=g+1}^m [k_{j0}^{(1)}\widetilde{V}_j^{(1)}(z)
+ k_{j0}^{(2)}\widetilde{V}_j^{(2)}(z)]\quad{\rm for~all}~~|z|>r_\#,
\eeq{2.32}
which proves that the coated cylinder and all the line dipoles inside the cloaking region
are invisible outside the cloaking region in this limit.

In this proof we have assumed that the dipole positions $z_j$ are independent of $\Gd$. If
they depend on $\Gd$ and $|z_i(\Gd)-z_j(\Gd)|$ is not bounded below by a positive constant
for all $i\ne j$ then it is an open question as to whether cloaking persists. At least in some
cases it may persist since \citeAPY{Nicorovici:2007:OCT} show that ``polarizable'' quadrupoles
are cloaked.

\section{Numerical examples of  cloaking of collections of polarizable line dipoles}

Due to the mathematical equivalence between the analysis for the coated
cylinder in the cases $r_{\rm c}>r_{\rm s}$ and $r_{\rm c}<r_{\rm s}$,
we can use the same numerical tools here as were employed
in the paper (\citeAY{Nicorovici:2007:OCT}) to solve for the fields in the folded geometry.
Then we use the unfolding transformation \eq{0.6} to obtain results for the potential
in the unfolded geometry, where the permittivity in the shell is anisotropic (with a positive
definite imaginary part) and given by \eq{0.7}. We have prepared
three animations illustrating the cloaking action, one for a pair of polarizable dipoles in a uniform external field,
and two others for a set of six polarizable dipoles arranged on the vertices of a hexagon. (These animations
show snapshots of the potential distribution in space, for a sequence of equilibrium solutions, as discussed
by \citeAPY{Nicorovici:2007:OCT}).  We present here in Figs. 3 and 4
images from each animation.

\begin{figure}
\centering
{\resizebox{3.0in}{3.0in}
{\includegraphics{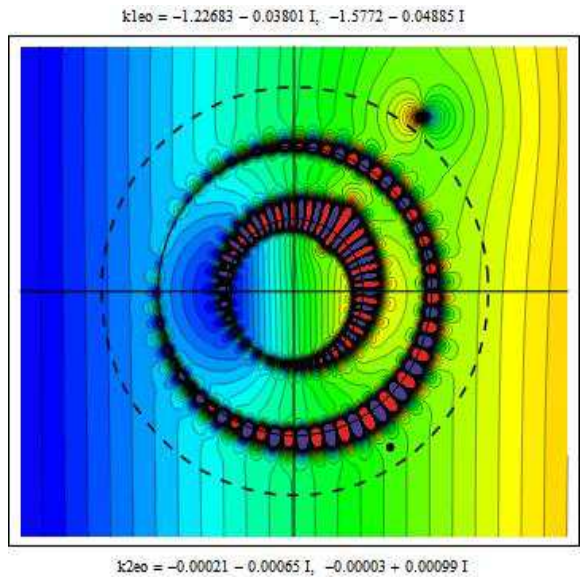}}}
\caption{Numerical computations for the potential associated with a pair of polarizable dipoles (polarizability
$\Ga=2$ located
at points $z_1=2.816-4.328 i$ and $z_2=3.755+4.828 i$
in the unfolded geometry (unfolding parameter $a=2$) with $\Gve_{\rm s}=-1+ 10^{-9} i$, $\Gve_{\rm c}=\Gve_{\rm m}=1$,
$r_{\rm c}=8$, $r_{\rm c}'=2$, $r_{\rm s}=r_{\rm s}'=4$.
The dashed line denotes the cloaking radius,  at $r_\#=\sqrt{r_{\rm c} r_{\rm s}}\simeq 5.657$. Note that one dipole is
outside the cloaking region, while the other is inside.}
\labfig{2}
\end{figure}

Fig.3 shows the potential associated with two polarizable dipoles, of which one is inside the cloaking radius and the other
outside it. The resonant region touches the cloaked line dipole, and quenches the field acting on it. As in the previous
study (\citeAY{Nicorovici:2007:OCT}), the resonance develops first on the shell-core boundary, before developing on the
shell-matrix boundary (see movie 1).

\begin{figure}
\vspace{1in}
\begin{minipage}[b]{3.0in}
{\resizebox{3.0in}{3.0in}
{\includegraphics{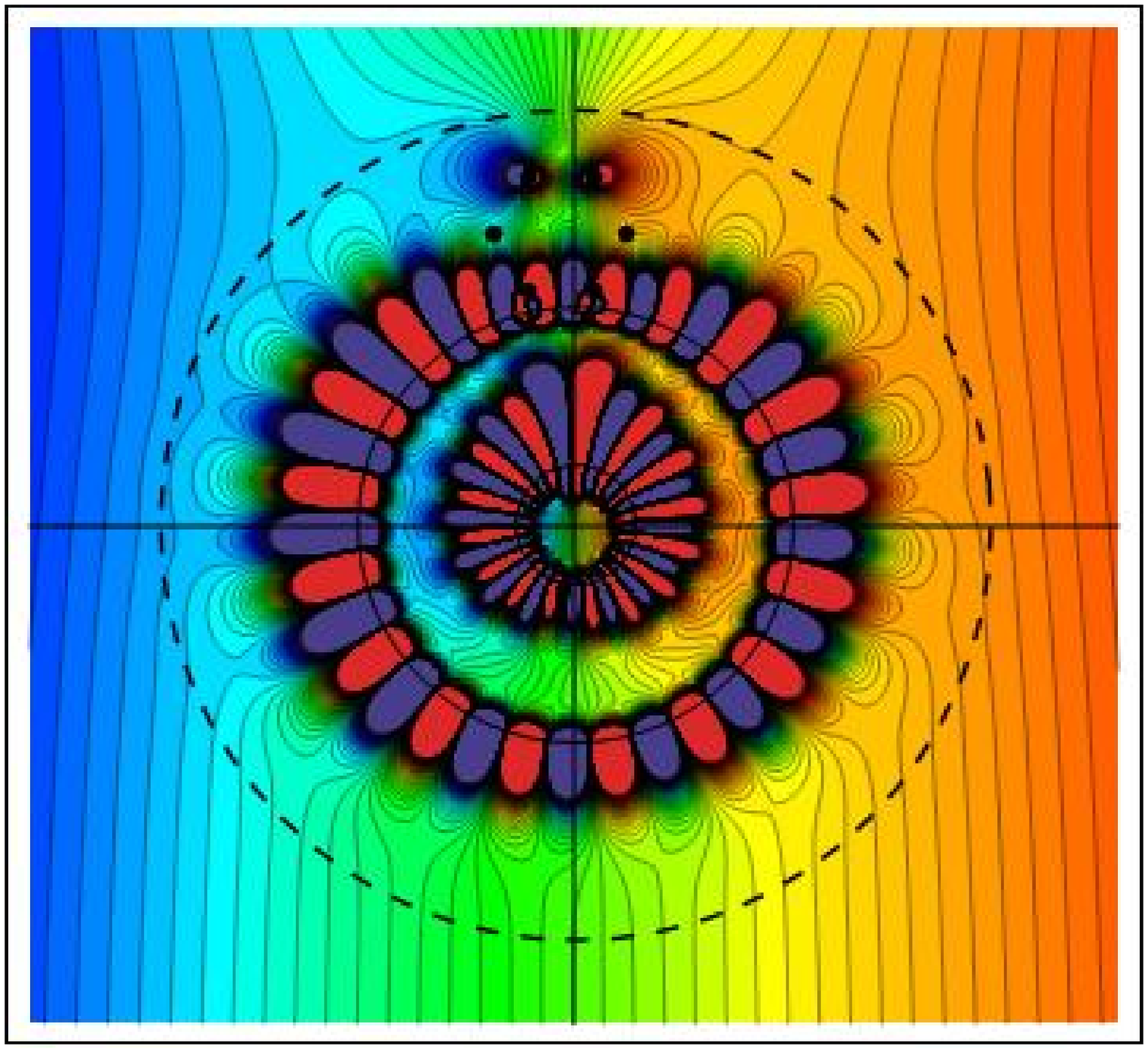}}}
\end{minipage}
~~
\begin{minipage}[b]{3.0in}
{\resizebox{3.0in}{3.0in}
{\includegraphics{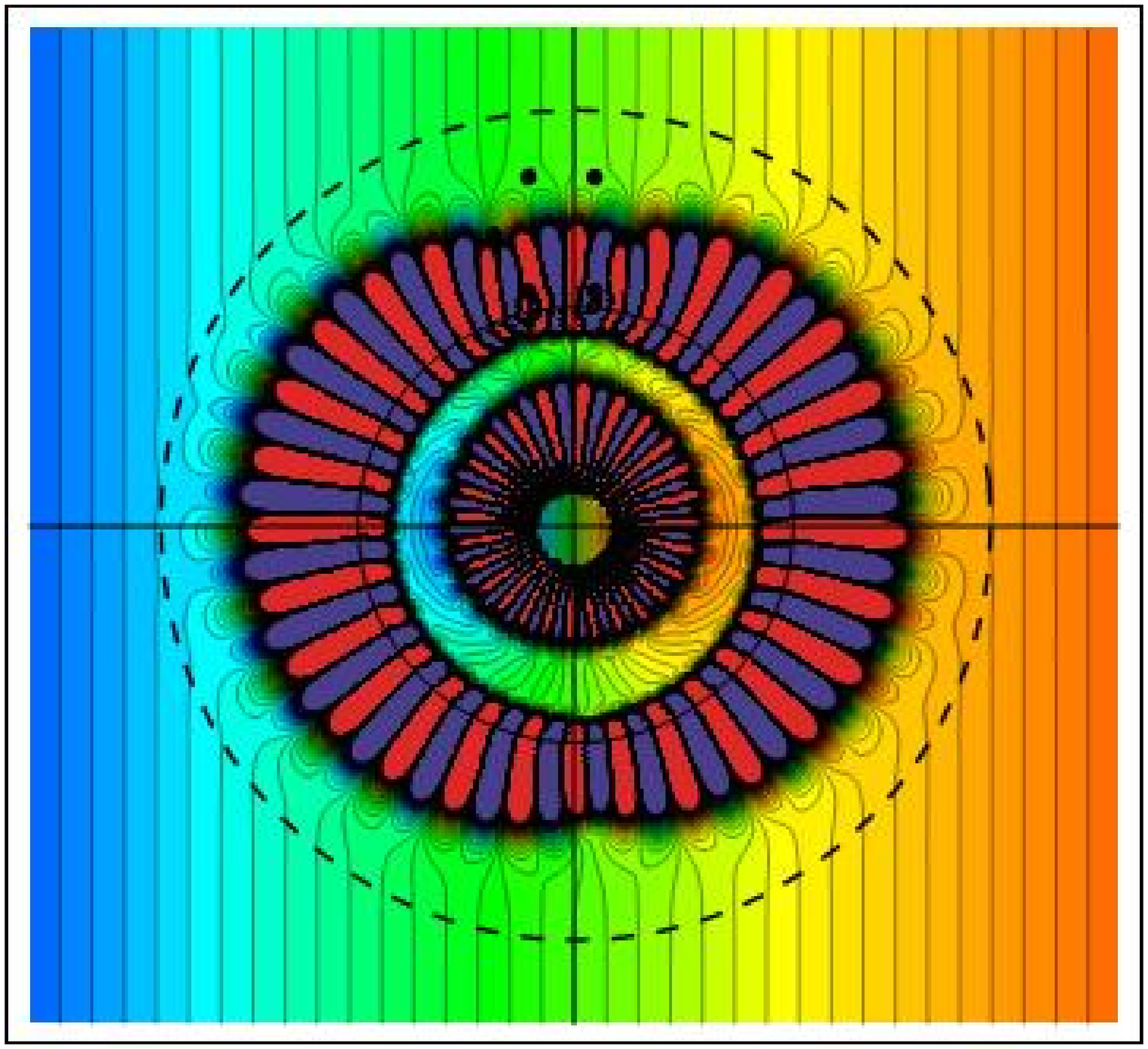}}}
\end{minipage}
\vspace{0.1in} \\
\begin{minipage}[b]{3.0in}
{\resizebox{3.0in}{0.3in}
{\includegraphics{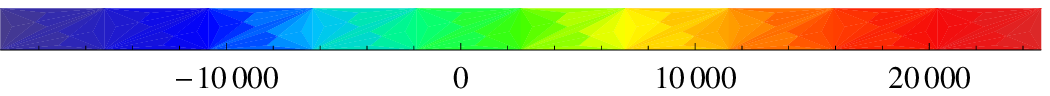}}}
\end{minipage}
~~
\begin{minipage}[b]{3.0in}
{\resizebox{3.0in}{0.3in}
{\includegraphics{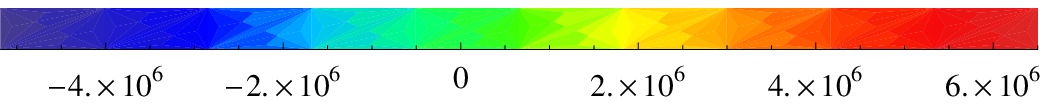}}}
\end{minipage}
\vspace{0.05in}
\caption{Numerical computations for the potential associated with a six polarizable dipoles arranged on the vertices
of a hexagon
in the unfolded geometry (unfolding parameter $a=r_{\rm s}'/r_{\rm c}'$) with (a) $\Gve_{\rm s}=-1+ 10^{-9} i$
and (b)$\Gve_{\rm s}=-1+10^{-15} i$. In both cases, $\Gve_{\rm m}=\Gve_{\rm c}=1$,
$r_{\rm c}=14.5455$, $r_{\rm c}'=1.1$, $r_{\rm s}=r_{\rm s}'=4$, while each line dipole has polarizability $\Ga=2$.
The dashed line denotes the cloaking radius, at $r_\#=\sqrt{r_{\rm c} r_{\rm s}}\simeq 7.6277$.}
\labfig{3}
\end{figure}

Fig.4 shows two frames from movies 2 and 3, and compares the cloaking of a set of six polarizable dipoles for two values of the imaginary part of $\Gve_{\rm s}$. As can be seen
from the first figure, an imaginary part of $10^{-9}$ is not sufficient to ensure cloaking of the two dipoles closest to $r_\#$.
However, good cloaking of all six dipoles is achieved for an imaginary part of $10^{-15}$. As in the papers
of \citeAPY{Bruno:2007:SCS} and
\citeAPY{Nicorovici:2007:OCT},
it appears that cloaking becomes more difficult as the number of polarizable particles in the collection increases, and becomes
more effective
as the particles move more deeply into the cloaking region.

\section*{Acknowledgements}
G.W.M is grateful to the National Science Foundation for support under grant DMS-070978.
The work of N.A.N. and R.McP. was supported by an Australian Research Council Discovery Grant.
Z.J. is thankful to the Army Research Office for support under the ARO-MURI award 50342-PH-MUR.
We are grateful to the referees for their comments on the manuscript.
\bibliography{/u/ma/milton/tcbook,/u/ma/milton/newref}


\ifx \bblindex \undefined \def \bblindex #1{} \fi\ifx \bblindex \undefined \def
  \bblindex #1{} \fi
\begin{thebibliography}{}

\ifx \xbblversion \undefined \input xbbl.sty \fi

\bibitem[\protect\citeauthoryear{Boardman and Marinov}{Boardman and
  Marinov}{2006}]{Boardman:2006:NRC}
\bblauthor{Boardman, A.~D.} and \bblauthor{K.~Marinov} \bblyear{2006}.
\newblock \bbltitle{Non-radiating and radiating configurations driven by
  left-handed metamaterials}.
\newblock {\em \bbljournal{Journal of the Optical Society of America B}\/}
  \bblvolume{23}\penalty0 (\bblnumber{3}):\penalty0 \bblpages{543--552}.
\showEXTRA{%
\showbibdate{\bblbibdate{Wed Aug 9 2006}}
}

\bibitem[\protect\citeauthoryear{Bruno and Lintner}{Bruno and
  Lintner}{2007}]{Bruno:2007:SCS}
\bblauthor{Bruno, O.~P.} and \bblauthor{S.~Lintner} \bblyear{2007}.
\newblock \bbltitle{Superlens-cloaking of small dielectric bodies in the
  quasistatic regime}.
\newblock {\em \bbljournal{Journal of Applied Physics}\/}
  \bblvolume{102}:\penalty0 \bblpages{124502}.
\showEXTRA{%
\showbibdate{\bblbibdate{Fri Mar 21 2008}}
}

\bibitem[\protect\citeauthoryear{Cui, Cheng, Lu, Jiang, and Kong}{Cui
  et~al.}{2005}]{Cui:2005:LEE}
\bblauthor{Cui, T.~J.}, \bblauthor{Q.~Cheng}, \bblauthor{W.~B. Lu},
  \bblauthor{Q.~Jiang}, and \bblauthor{J.~A. Kong} \bblyear{2005}.
\newblock \bbltitle{Localization of electromagnetic energy using a
  left-handed-medium slab}.
\newblock {\em \bbljournal{Physical Review B (Solid State)}\/}
  \bblvolume{71}:\penalty0 \bblpages{045114}.
\showEXTRA{%
\showCODEN{\bblCODEN{PRBMDO}}
\showISSN{\bblISSN{0163-1829}}
\showbibdate{\bblbibdate{Wed Dec 14 2005}}
}

\bibitem[\protect\citeauthoryear{Jacob, Alekseyev, and Narimanov}{Jacob
  et~al.}{2006}]{Jacob:2006:OHF}
\bblauthor{Jacob, Z.}, \bblauthor{L.~V. Alekseyev}, and
  \bblauthor{E.~Narimanov} \bblyear{2006}.
\newblock \bbltitle{Optical hyperlens: Far-field imaging beyond the diffraction
  limit}.
\newblock {\em \bbljournal{Optics Express}\/} \bblvolume{14}\penalty0
  (\bblnumber{18}):\penalty0 \bblpages{8247--8256}.
\showEXTRA{%
\showbibdate{\bblbibdate{Sat Apr 19 2008}}
}

\bibitem[\protect\citeauthoryear{Kildishev and Narimanov}{Kildishev and
  Narimanov}{2007}]{Kildishev:2007:IMH}
\bblauthor{Kildishev, A.~V.} and \bblauthor{E.~E. Narimanov} \bblyear{2007}.
\newblock \bbltitle{Impedance-matched hyperlens}.
\newblock {\em \bbljournal{Optics Letters}\/} \bblvolume{32}\penalty0
  (\bblnumber{23}):\penalty0 \bblpages{3432--3434}.
\showEXTRA{%
\showbibdate{\bblbibdate{Fri Mar 21 2008}}
}

\bibitem[\protect\citeauthoryear{Leonhardt and Philbin}{Leonhardt and
  Philbin}{2006}]{Leonhardt:2006:GRE}
\bblauthor{Leonhardt, U.} and \bblauthor{T.~G. Philbin} \bblyear{2006}.
\newblock \bbltitle{General relativity in electrical engineering}.
\newblock {\em \bbljournal{New Journal of Physics}\/} \bblvolume{8}:\penalty0
  \bblpages{247}.
\showEXTRA{%
\showbibdate{\bblbibdate{Wed Aug 23 2006}}
}

\bibitem[\protect\citeauthoryear{Milton and Nicorovici}{Milton and
  Nicorovici}{2006}]{Milton:2006:CEA}
\bblauthor{Milton, G.~W.} and \bblauthor{N.-A.~P. Nicorovici} \bblyear{2006}.
\newblock \bbltitle{On the cloaking effects associated with anomalous localized
  resonance}.
\newblock {\em \bbljournal{Proceedings of the Royal Society of London. Series
  A, Mathematical and Physical Sciences}\/} \bblvolume{462}\penalty0
  (\bblnumber{2074}):\penalty0 \bblpages{3027--3059}.
\newblock \bblnote{Published online May 3rd: doi:10.1098/rspa.2006.1715}.
\showEXTRA{%
\showCODEN{\bblCODEN{PRLAAZ}}
\showISSN{\bblISSN{0080-4630}}
\showbibdate{\bblbibdate{Mon Jul 3 2006}}
}

\bibitem[\protect\citeauthoryear{Milton, Nicorovici, and McPhedran}{Milton
  et~al.}{2007}]{Milton:2006:OPL}
\bblauthor{Milton, G.~W.}, \bblauthor{N.-A.~P. Nicorovici}, and
  \bblauthor{R.~C. McPhedran} \bblyear{2007}.
\newblock \bbltitle{Opaque perfect lenses}.
\newblock {\em \bbljournal{Physica B}\/} \bblvolume{394}:\penalty0
  \bblpages{171--175}.
\showEXTRA{%
\showbibdate{\bblbibdate{Fri Mar 21 2008}}
}

\bibitem[\protect\citeauthoryear{Milton, Nicorovici, McPhedran, and
  Podolskiy}{Milton et~al.}{2005}]{Milton:2005:PSQ}
\bblauthor{Milton, G.~W.}, \bblauthor{N.-A.~P. Nicorovici}, \bblauthor{R.~C.
  McPhedran}, and \bblauthor{V.~A. Podolskiy} \bblyear{2005}.
\newblock \bbltitle{A proof of superlensing in the quasistatic regime, and
  limitations of superlenses in this regime due to anomalous localized
  resonance}.
\newblock {\em \bbljournal{Proceedings of the Royal Society of London. Series
  A, Mathematical and Physical Sciences}\/} \bblvolume{461}:\penalty0
  \bblpages{3999--4034}.
\showEXTRA{%
\showCODEN{\bblCODEN{PRLAAZ}}
\showISSN{\bblISSN{0080-4630}}
\showbibdate{\bblbibdate{Mon Dec 12 2005}}
}

\bibitem[\protect\citeauthoryear{Nicorovici, McPhedran, and Milton}{Nicorovici
  et~al.}{1994}]{Nicorovici:1994:ODP}
\bblauthor{Nicorovici, N.~A.}, \bblauthor{R.~C. McPhedran}, and
  \bblauthor{G.~W. Milton} \bblyear{1994}.
\newblock \bbltitle{Optical and dielectric properties of partially resonant
  composites}.
\newblock {\em \bbljournal{Physical Review B (Solid State)}\/}
  \bblvolume{49}\penalty0 (\bblnumber{12}):\penalty0 \bblpages{8479--8482}.
\showEXTRA{%
\showCODEN{\bblCODEN{PRBMDO}}
\showISSN{\bblISSN{0163-1829}}
\showbibdate{\bblbibdate{Fri Nov 5 13:45:09 MST 1999}}
}

\bibitem[\protect\citeauthoryear{Nicorovici, Milton, McPhedran, and
  Botten}{Nicorovici et~al.}{2007}]{Nicorovici:2007:OCT}
\bblauthor{Nicorovici, N.-A.~P.}, \bblauthor{G.~W. Milton}, \bblauthor{R.~C.
  McPhedran}, and \bblauthor{L.~C. Botten} \bblyear{2007}.
\newblock \bbltitle{Quasistatic cloaking of two-dimensional polarizable
  discrete systems by anomalous resonance}.
\newblock {\em \bbljournal{Optics Express}\/} \bblvolume{15}\penalty0
  (\bblnumber{10}):\penalty0 \bblpages{6314--6323}.
\showEXTRA{%
\showbibdate{\bblbibdate{Tue Jun 12 2007}}
}

\bibitem[\protect\citeauthoryear{Pendry}{Pendry}{2000}]{Pendry:2000:NRM}
\bblauthor{Pendry, J.~B.} \bblyear{2000}.
\newblock \bbltitle{Negative refraction makes a perfect lens}.
\newblock {\em \bbljournal{Physical Review Letters}\/} \bblvolume{85}:\penalty0
  \bblpages{3966--3969}.
\showEXTRA{%
\showbibdate{\bblbibdate{Mon May 30 2005}}
}

\bibitem[\protect\citeauthoryear{Rahm, Schurig, Roberts, Cummer, Smith, and
  Pendry}{Rahm et~al.}{2008}]{Rahm:2008:DEC}
\bblauthor{Rahm, M.}, \bblauthor{D.~Schurig}, \bblauthor{D.~A. Roberts},
  \bblauthor{S.~A. Cummer}, \bblauthor{D.~R. Smith}, and \bblauthor{J.~B.
  Pendry} \bblyear{2008}.
\newblock \bbltitle{Design of electromagnetic cloaks and concentrators using
  form-invariant coordinate transformations of maxwell's equations}.
\newblock {\em \bbljournal{Photonics and Nanostructures -- Fundamentals and
  Applications}\/} \bblvolume{6}:\penalty0 \bblpages{87--95}.
\showEXTRA{%
\showbibdate{\bblbibdate{Sat Apr 19 2008}}
}

\bibitem[\protect\citeauthoryear{Salandrino and Engheta}{Salandrino and
  Engheta}{2006}]{Salandrino:2006:FFS}
\bblauthor{Salandrino, A.} and \bblauthor{N.~Engheta} \bblyear{2006}.
\newblock \bbltitle{Far-field subdiffraction optical microscopy using
  metamaterial crystals: Theory and simulations}.
\newblock {\em \bbljournal{Physical Review B (Solid State)}\/}
  \bblvolume{74}:\penalty0 \bblpages{075103}.
\showEXTRA{%
\showbibdate{\bblbibdate{Sat Jun 11 2005}}
}

\bibitem[\protect\citeauthoryear{Veselago}{Veselago}{1967}]{Veselago:1967:ESS}
\bblauthor{Veselago, V.~G.} \bblyear{1967}.
\newblock \bbltitle{The electrodynamics of substances with simultaneously
  negative values of $\epsilon$ and $\mu$}.
\newblock {\em \bbljournal{Uspekhi Fizicheskikh Nauk}\/}
  \bblvolume{92}:\penalty0 \bblpages{517--526}.
\newblock \bblnote{English translation in {\em Soviet Physics Uspekhi}
  10:509--514 (1968)}.
\showEXTRA{%
\showCODEN{\bblCODEN{UFNAAG}}
\showISSN{\bblISSN{0042-1294}}
\showbibdate{\bblbibdate{Thu May 2 13:45:09 MST 2005}}
}

\end{thebibliography}

\end{document}